%% file: main.tex
\documentclass[sigconf,nonacm]{acmart}

\usepackage{subfigure}
\usepackage{siunitx}
\usepackage{enumitem}

\usepackage{algorithm}
\usepackage[noend]{algpseudocode}
\let\ReturnInline\Return
\renewcommand{\Return}{\State\ReturnInline}
\algrenewcommand\algorithmicrequire{$\rhd$}
\algrenewcommand\algorithmicensure{$\square$}

\AtBeginDocument{%
  \providecommand\BibTeX{{%
    \normalfont B\kern-0.5em{\scshape i\kern-0.25em b}\kern-0.8em\TeX}}}

\setcopyright{acmcopyright}
\copyrightyear{2018}
\acmYear{2018}
\acmDOI{XXXXXXX.XXXXXXX}

\acmConference[Conference acronym 'XX]{Make sure to enter the correct
  conference title from your rights confirmation emai}{June 03--05,
  2018}{Woodstock, NY}

\newcommand{\ignore}[1]{}

\begin{document}

\title[A Starting Point for Dynamic Community Detection with Leiden Algorithm]{A Starting Point for Dynamic Community Detection\\ with Leiden Algorithm}


\author{Subhajit Sahu}
\email{subhajit.sahu@research.iiit.ac.in}
\affiliation{%
  \institution{IIIT Hyderabad}
  \streetaddress{Professor CR Rao Rd, Gachibowli}
  \city{Hyderabad}
  \state{Telangana}
  \country{India}
  \postcode{500032}
}


\settopmatter{printfolios=true}

\begin{abstract}
Real-world graphs often evolve over time, making community or cluster detection a crucial task. In this technical report, we extend three dynamic approaches --- Naive-dynamic (ND), Delta-screening (DS), and Dynamic Frontier (DF) --- to our multicore implementation of the Leiden algorithm, known for its high-quality community detection. Our experiments, conducted on a server with a 64-core AMD EPYC-7742 processor, show that ND, DS, and DF Leiden achieve average speedups of $1.37\times$, $1.47\times$, and $1.98\times$ on large graphs with random batch updates\ignore{, and $1.07\times$, $1.10\times$, and $1.13\times$ on real-world dynamic graphs}, compared to the Static Leiden algorithm --- while scaling at a rate of $1.6\times$ for every doubling of threads. To our knowledge, this is the first attempt to apply dynamic approaches to the Leiden algorithm. We hope these early results pave the way for further development of dynamic approaches for evolving graphs.
\end{abstract}



\maketitle

\section{Introduction}
\label{sec:introduction}
\input{01-introduction}

\section{Related work}
\label{sec:related}
\input{02-related-work}

\section{Preliminaries}
\label{sec:preliminaries}
\input{03-preliminaries}

\section{Approach}
\label{sec:approach}
\input{04-approach}
\section{Evaluation}
\label{sec:evaluation}
\input{05-evaluation}

\section{Conclusion}
\label{sec:conclusion}
\input{06-conclusion}

\begin{acks}
I would like to thank Prof. Kishore Kothapalli and Prof. Dip Sankar Banerjee for their support.
\end{acks}

\bibliographystyle{ACM-Reference-Format}
\bibliography{main}

\clearpage
\appendix
\section{Appendix}
\input{aa-appendix}

\end{document}

%% file: 01-introduction.tex
Driven by the ability of graphs to represent complex real-world data and capture intricate relationships among entities, research in graph-structured data has been rapidly growing. A core focus of this field is community detection, which involves decomposing a graph into densely connected groups --- revealing the natural structure inherent in the data. Discovering hidden communities in social networks \cite{blekanov2021detection}, analyzing regional retail landscapes \cite{verhetsel2022regional}, studying user relations in Decentralized Online Social Networks (DOSNs) \cite{la2022information}, delineating health care service areas \cite{wang2021network}, uncovering disinformation networks in Telegram \cite{la2021uncovering}, partitioning large graphs for machine learning \cite{bai2024leiden}, studying biological processes \cite{heumos2023best}\ignore{\cite{liu2024sclega, hartman2024peptide, muller2024spatialleiden}}, characterizing polarized information ecosystems\ignore{(climate change conversations)} \cite{uyheng2021mainstream}, developing cyber resilient systems\ignore{/networks by investigating cyber defense techniques} \cite{chernikova2022cyber}, identifying transportation patterns \cite{chen2023deciphering}, identifying attacks in blockchain networks \cite{erfan2023community}, examining restored accounts on Twitter \cite{kapoor2021ll}, exploring the eco-epidemiology of zoonoses \cite{desvars2024one}, analyzing linguistic variation in memes \cite{zhou2023social}, reconstructing multi-step cyber attacks \cite{zang2023attack}, studying twitter communities during the 2022 war in Ukraine \cite{sliwa2024case}, and automated microservice decomposition \cite{cao2022implementation} are all applications of community detection.

A challenge in community detection is the absence of prior knowledge about the number and size distribution of communities. To address this, researchers have developed numerous heuristics for finding communities \cite{com-blondel08, com-gregory10}\ignore{\cite{com-raghavan07, com-guimera05, com-derenyi05, com-newman06, com-reichardt06, com-rosvall08, infomap-rosvall09, com-fortunato10, com-kloster14, com-come15, com-ruan15, com-newman16, com-ghoshal19, com-rita20, com-lu20, com-gupta22}}. The quality of identified communities is often measured using fitness metrics such as the modularity score proposed by Newman et al. \cite{com-newman04}. These communities are\ignore{considered} intrinsic when identified based solely on network topology\ignore{, without external attributes}, and they are disjoint when each vertex belongs to only one community \cite{com-gregory10}.

The Louvain method, proposed by Blondel et al. \cite{com-blondel08}\ignore{from the University of Louvain}, is one of the most popular community detection algorithms \cite{com-lancichinetti09}. This greedy algorithm employs a two-phase approach, consisting of an iterative local-moving phase and an aggregation phase, to iteratively optimize the modularity metric over several passes \cite{com-blondel08}. It has a time complexity of $O(LM)$ (where $M$ is the number of edges in the graph and $L$ is the total number of iterations performed across all passes) and efficiently identifies communities with high modularity.\ignore{What are the popular CD algorithms?}

Despite its popularity, the Louvain method has been observed to produce internally disconnected and poorly connected communities. This problem occurs during the local-moving phase, where important nodes ---\ignore{such as hubs} that serve as bridges within a community --- may be reassigned to another community to which they have stronger connections, breaking the connectivity of their original community. To address these issues, Traag et al. \cite{com-traag19}\ignore{from the University of Leiden} proposed the \textbf{Leiden algorithm}\ignore{ (which is partially based on the smart local move algorithm \cite{waltman2013smart})}, which introduces a refinement phase between the local-moving and aggregation phases. During the refinement phase, vertices can explore and potentially form sub-communities within the communities identified in the local-moving phase --- allowing the algorithm to identify well-connected communities \cite{com-traag19}.

But many real-world graphs are immense and evolve rapidly over time\ignore{, through the insertion and deletion of edges and vertices}. For efficiency, algorithms are needed that update results without recomputing from scratch, known as \textbf{dynamic algorithms}. Dynamic community detection algorithms also allow one to track the evolution of communities over time, identifying key events like growth, shrinkage, merging, splitting, birth, and death. However, research efforts have focused on detecting communities in dynamic networks using the Louvain algorithm. None of the works have extended these approaches to the Leiden algorithm.

This technical report extends the Naive-dynamic (ND) \cite{com-aynaud10}\ignore{\cite{com-chong13, com-shang14, com-zhuang19}}, Delta-screening (DS) \cite{com-zarayeneh21}, and the recently proposed parallel Dynamic Frontier (DF) approach \cite{sahu2024shared} to the Leiden algorithm.\footnote{\url{https://github.com/puzzlef/leiden-communities-openmp-dynamic}} Our algorithms build on top of GVE-Leiden \cite{sahu2024fast}, one of the most efficient shared-memory implementations of the Static Leiden algorithm.\ignore{Upon receiving a batch update comprising edge deletions and insertions, DF Leiden incrementally identifies and processes an approximate set of affected vertices in an incremental manner. This work represents, to the best of our knowledge, the first endeavor in extending existing dynamic approaches to the Leiden algorithm.}

\subsection{Our Contributions}

We propose the first dynamic algorithms for Leiden, extending dynamic approaches originally developed for Louvain \cite{com-aynaud10, com-zarayeneh21, sahu2024shared} to Leiden. Our contributions are as follows:

\begin{itemize}
  \item Directly applying DF approach \cite{sahu2024shared} to Leiden fails: On a small batch update Leiden's local-moving phase converges quickly. Then refinement phase, divides communities into smaller sub-communities. Terminating at this point (as DF Louvain) results in poor modularity, as these sub-communities require hierarchical aggregation. Consequently, multiple passes of Leiden are needed despite early convergence. This is a key insight.
  \item For selective refinement we track vertices that migrate between communities during local-moving phase, and marking both source and target communities for refinement. Untouched communities are not refined. This reduces the processing costs, as refined communities must be processed in the subsequent passes.
  \item Refinement begins by initializing each vertex as its own sub-community. These merge into larger sub-communities within their own community bounds. However, this approach applied to a subset of communities can result in internal disconnection. This occurs because a vertex $i$ may not belong to community $C$ with the same ID (i.e., $C = i$), but the initialization places $i$ in an isolated sub-community, while other vertices in community $C$ may not be connected to $i$. As refinement continues, more vertices join $i$ in sub-community $D = i$, yet it remains disconnected from rest of community $C = i$. This insight is illustrated in Figure \ref{fig:subrefine-issue}, and explained in detail in Section \ref{sec:subset-refine-method}.
  \item To resolve this, we propose a subset renumbering procedure where, for each community $C$, the ID of any member vertex $i'$ is selected. This ID is then used to renumber the community membership of all vertices in $C$, update the total edge weight, and set the changed communities flags to the new ID $j$. Afterward, selective refinement can be performed.
  \item We also select communities for refinement based on the batch update (changed communities) - those with edge deletions or insertions within the community. Cross-community edge deletions and insertions are automatically selected when vertices migrate between communities.
  \item Selective refinement causes load imbalance during aggregation phase, as unrefined communities remain large while refined ones are small. To address this, we reduce chunk size for OpenMP parallelization during aggregation, which introduces scheduling overhead but ensures load balancing.
  \item We present novel algorithms\ignore{in this report}, including methods for marking changed-communities and subset-renumbering.
\end{itemize}

\noindent
While reported speedups are modest (DF Leiden is $3.7\times$ faster than Static Leiden for small updates), we believe these insights are crucial for developing more efficient dynamic algorithms for Leiden.

%% file: 02-related-work.tex
A core idea for dynamic community detection, among most approaches, is to use the community membership of each vertex from the previous snapshot of the graph, instead of initializing each vertex into singleton communities. One straightforward strategy for dynamic community detection involves leveraging the community memberships of vertices from the previous snapshot of the graph \cite{com-aynaud10, com-chong13, com-shang14, com-zhuang19}, which has been termed as \textbf{Naive-dynamic (ND)}. Alternatively, more sophisticated techniques have been devised to reduce computational overhead by identifying a smaller subset of the graph affected by changes. These techniques include updating only changed vertices \cite{com-aktunc15, com-yin16}, processing vertices in the proximity of updated edges (within a specified threshold distance) \cite{com-held16}, disbanding affected communities into lower-level networks \cite{com-cordeiro16}, or employing a dynamic modularity metric to recompute community memberships from scratch \cite{com-meng16}.

Zarayeneh et al. \cite{com-zarayeneh21} propose the \textbf{Delta-screening (DS)} approach for updating communities in a dynamic graph. This technique examines edge deletions and insertions to the original graph, and identifies a subset of vertices that are likely to be impacted by the change using the modularity objective. Subsequently, only the identified subsets are processed for community state updates, using the Louvain and Smart Local-Moving (SLM) algorithms \cite{com-waltman13}. Recently, Sahu et al. \cite{sahu2024shared} introduced the parallel \textbf{Dynamic Frontier (DF)} approach for the Louvain algorithm. Given a batch update consisting of edge deletions or edge insertions, the DF approach incrementally identifies an approximate set of affected vertices in the graph with a low run time overhead. They also discuss a parallel algorithm for the Delta-screening (DS) approach.

However, to the best of our knowledge, none of the proposed approaches have been extended to the Leiden algorithm.

%% file: 03-preliminaries.tex
Let $G(V, E, w)$ represent an undirected graph, where $V$ denotes the set of vertices, $E$ denotes the set of edges, and $w_{ij} = w_{ji}$ signifies a positive weight associated with each edge in the graph. In the case of an unweighted graph, we assume each edge has a unit weight ($w_{ij} = 1$). Additionally, we denote the neighbors of each vertex $i$ as $J_i = \{j\ |\ (i, j) \in E\}$, the weighted degree of each vertex $i$ as $K_i = \sum_{j \in J_i} w_{ij}$, the total number of vertices in the graph as $N = |V|$, the total number of edges in the graph as $M = |E|$, and the sum of edge weights in the undirected graph as $m = \sum_{i, j \in V} w_{ij}/2$.

\subsection{Community detection}

Disjoint community detection involves finding a community membership mapping, $C: V \rightarrow \Gamma$, which maps each vertex $i \in V$ to a community ID $c \in \Gamma$, where $\Gamma$ is the set of community-ids. The vertices within a community $c$ are denoted as $V_c$, and the community to which a vertex $i$ belongs is denoted as $C_i$. Furthermore, we define the neighbors of vertex $i$ belonging to a community $c$ as $J_{i \rightarrow c} = \{j\ |\ j \in J_i\ and\ C_j = c\}$, the sum of those edge weights as $K_{i \rightarrow c} = \{w_{ij}\ |\ j \in J_{i \rightarrow c}\}$, the sum of edge weights within $c$ as $\sigma_c = \sum_{(i, j) \in E\ and\ C_i = C_j = c} w_{ij}$, and the total edge weight of $c$ as $\Sigma_c = \sum_{(i, j) \in E\ and\ C_i = c} w_{ij}$ \cite{com-zarayeneh21}.

\subsection{Modularity}

Modularity is a metric used to assess the quality of communities identified by community detection algorithms, which are\ignore{typically} heuristic-based \cite{com-newman04}. It ranges from $[-0.5, 1]$ (higher is better), and is calculated as the difference between the fraction of edges within communities and the expected fraction of edges if they were distributed randomly \cite{com-brandes07}.\ignore{Theoretically, optimizing this function leads to the best possible grouping\ignore{ of the network} \cite{com-newman04, com-traag11}.} The modularity $Q$ of the obtained communities can be calculated using Equation \ref{eq:modularity}\ignore{, where $\delta$ represents the Kronecker delta function ($\delta (x,y)=1$ if $x=y$ and $0$ otherwise)}. The \textit{delta modularity} of moving a vertex $i$ from community $d$ to community $c$, denoted as $\Delta Q_{i: d \rightarrow c}$, is determined using Equation \ref{eq:delta-modularity}.

\begin{equation}
\label{eq:modularity}
  Q
  = \sum_{c \in \Gamma} \left[\frac{\sigma_c}{2m} - \left(\frac{\Sigma_c}{2m}\right)^2\right]
\end{equation}

\begin{equation}
\label{eq:delta-modularity}
  \Delta Q_{i: d \rightarrow c}
  = \frac{1}{m} (K_{i \rightarrow c} - K_{i \rightarrow d}) - \frac{K_i}{2m^2} (K_i + \Sigma_c - \Sigma_d)
\end{equation}

\subsection{Louvain algorithm}
\label{sec:about-Louvain}

The Louvain method is a greedy, modularity optimization based agglomerative algorithm for detecting communities of high quality in a graph. It works in two phases: in the local-moving phase, each vertex $i$ considers moving to a neighboring community $\{C_j\ |\ j \in J_i\}$ to maximize modularity increase $\Delta Q_{i:C_i \rightarrow C_j}$. In the aggregation phase, vertices in the same community are merged into super-vertices. These phases constitute a pass, and repeat until modularity gain stops. This produces a hierarchical structure (dendrogram), with the top level representing the final communities.

\subsection{Leiden algorithm}
\label{sec:about-leiden}

As previously mentioned, the Louvain method, while effective, can produce internally disconnected or poorly connected communities. To address these issues, Traag et al. \cite{com-traag19} proposed the Leiden algorithm. This algorithm introduces a \textit{refinement phase} following the local-moving phase. During this phase, vertices within each community undergo constrained merges into other sub-communities within their community bounds (established during the local moving phase), starting from a singleton sub-community. This process is randomized, with the probability of a vertex joining a neighboring sub-community being proportional to the delta-modularity of the move. This step helps identify sub-communities within those identified during the local-moving phase. The Leiden algorithm ensures that all communities are well-separated and well-connected. Once the communities have converged, it guarantees that all vertices are optimally assigned and all communities are subset optimal \cite{com-traag19}. The Leiden algorithm has a time complexity of $O(L|E|)$, where $L$ is the total number of iterations performed, and a space complexity of $O(|V| + |E|)$, similar to the Louvain method.

\subsection{Dynamic approaches}
\label{sec:dynamic-graphs}

A dynamic graph can be represented as a sequence of graphs, where $G^t(V^t, E^t, w^t)$ denotes the graph at time step $t$. The changes between the graphs $G^{t-1}(V^{t-1}, E^{t-1}, w^{t-1})$ and $G^t(V^t, E^t, w^t)$ at consecutive time steps $t-1$ and $t$ can be described as a batch update $\Delta^t$ at time step $t$. This batch update consists of a set of edge deletions $\Delta^{t-} = \{(i, j)\ |\ i, j \in V\} = E^{t-1} \setminus E^t$ and a set of edge insertions $\Delta^{t+} = \{(i, j, w_{ij})\ |\ i, j \in V; w_{ij} > 0\} = E^t \setminus E^{t-1}$ \cite{com-zarayeneh21}. We refer to the scenario where $\Delta^t$ includes multiple edges being deleted and inserted as a \textit{batch update}.

\subsubsection{Naive-dynamic (ND) approach \cite{com-aynaud10, com-chong13, com-shang14, com-zhuang19}}
\label{sec:naive-dynamic}

The Naive dynamic approach\ignore{, presented by Aynaud et al. \cite{com-aynaud10},} is a simple method for detecting communities in dynamic networks. In this approach, vertices are assigned to communities based on the previous snapshot of the graph, and all vertices are processed, regardless of the edge deletions and insertions in the batch update (hence the term \textit{naive}).

\subsubsection{Delta-screening (DS) approach \cite{com-zarayeneh21}}
\label{sec:delta-screening}

This is a dynamic community detection approach that utilizes modularity-based scoring to identify an approximate region of the graph where vertices are likely to alter their community membership. In the DS approach, Zarayeneh et al. first sort the batch update consisting of edge deletions $(i, j) \in \Delta^{t-}$ and insertions $(i, j, w) \in \Delta^{t+}$ by their source vertex ID. For edge deletions within the same community, they mark $i$'s neighbors and $j$'s community as affected. For edge insertions across communities, they select the vertex $j^*$ with the highest modularity change among all insertions linked to vertex $i$ and mark $i$'s neighbors and $j^*$'s community as affected. Edge deletions between different communities and edge insertions within the same community are unlikely to impact community membership and are therefore ignored. Figure \ref{fig:about-cases--delta} shows the vertices (and communities) connected to a single source vertex $i$ that are identified as affected by the DS approach, in response to a batch update with both edge deletions and insertions.

\subsubsection{Dynamic Frontier (DF) approach \cite{sahu2024shared}}
\label{sec:about-frontier}

This approach first initializes each vertex's community membership to that obtained in the previous graph snapshot. In instances where edge deletions occur between vertices within the same community or edge insertions between vertices in different communities, the DF approach marks the source vertex $i$ as affected\ignore{, as depicted by the yellow-highlighted vertices in Figure \ref{fig:about-cases--frontier}}. As batch updates are undirected, both endpoints $i$ and $j$ are effectively marked as affected. Edge deletions spanning different communities or edge insertions within the same community are disregarded (as previously mentioned in Section \ref{sec:delta-screening}). Furthermore, when a vertex $i$ alters its community membership during the community detection process\ignore{(illustrated by $i$ transitioning from its original community in the center to its new community on the right)}, all its neighboring vertices $j \in J_i$ are marked as affected\ignore{, as depicted in Figure \ref{fig:about-cases--frontier} (highlighted in yellow)}, while $i$ is marked as unaffected. To minimize unnecessary computation, an affected vertex $i$ is also marked as unaffected even if its community remains unchanged. Thus, the DF approach follows a graph traversal-like process until the vertices' community assignments converge. Figure \ref{fig:about-cases--frontier} shows vertices connected to a single source vertex $i$ identified as affected by the DF approach in response to a batch update.

\input{src/tab-terminology}
\input{src/fig-about-cases}

%% file: src/tab-terminology.tex
\begin{table}[hbtp]
  \centering
  \caption{Key terminology used in this report.}
  \label{tab:terminology}
  \begin{tabular}{|c|p{39ex}|}
    \toprule
    \textbf{Term} &
    \textbf{Definition} \\
    \midrule
    Phase & Steps in Leiden algorithm: local-moving, refinement, and aggregation. \\ \hline
    Pass & One full cycle of the three phases. Multiple passes are repeated until convergence. \\ \hline
    Community bound & Community structure identified during the local-moving phase. \\ \hline
    Sub-community & Smaller groups formed during the refinement phase within community bounds. \\ \hline
  \bottomrule
  \end{tabular}
\end{table}

%% file: src/fig-about-cases.tex
\begin{figure}[hbtp]
  \centering
  \subfigure[Delta-screening (DS)]{
    \label{fig:about-cases--delta}
    \includegraphics[width=0.468\linewidth]{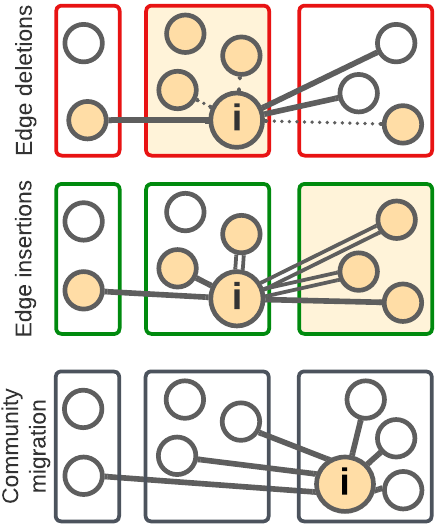}
  }
  \subfigure[Dynamic Frontier (DF)]{
    \label{fig:about-cases--frontier}
    \includegraphics[width=0.43\linewidth]{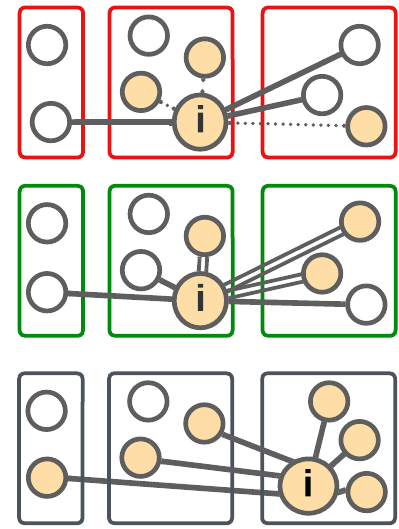}
  } \\[-1ex]
  \caption{Illustration of \textit{Delta-screening (DS)} \cite{com-zarayeneh21} and \textit{Dynamic Frontier (DF)} approaches \cite{sahu2024shared}, in the presence of edge deletions and insertions, represented with dotted lines and doubled lines, respectively. Vertices identified as affected (initial) by each approach are highlighted in brown, and entire communities marked as affected are depicted in light brown.}
  \label{fig:about-cases}
\end{figure}

%% file: 04-approach.tex
We observe that Dynamic Frontier (DF) Leiden outperforms both Naive-dynamic (ND) and Delta-screening (DS) Leiden, as shown in Section \ref{sec:performance-comparison}. Therefore, we focus most of our discussion below on how to extend the DF approach to the Leiden algorithm. The procedure to extend the ND and DS approaches to the Leiden algorithm are quite similar\ignore{, and are discussed at the end of this section}. Table \ref{tab:terminology} shows the terminology used in this report.

\subsection{No continued passes}
\label{sec:no-continued-passes}

A straightforward application of the DF approach to the Leiden algorithm involves processing incrementally identified affected vertices during the local-moving phase of the algorithm, refining the obtained communities in the subsequent phase, aggregating the refined communities into a super-vertex graph (where each refined community is collapsed into a super-vertex), and repeating this process until convergence is reached. However, for small batch updates, convergence may occur after just one pass of the algorithm, at which point the algorithm would terminate. This results in suboptimal communities with low modularity, as the communities generated from the refinement phase did not have the opportunity to hierarchically merge and form tightly-knit groups with dense internal connections and sparse inter-community links. An example of the communities returned in the absence of further processing is shown in Figure \ref{fig:subrefine-stages--01}, given a batch update. It is important to note that this issue does not arise in the absence of the refinement phase --- it occurs specifically when refinement is applied, as it forces the communities identified during the local-moving phase to be divided into smaller sub-communities. 

\input{src/fig-subrefine-stages}

\subsection{Full Refine method}
\label{sec:full-refine-method}

Despite the above mentioned issues, we\ignore{still} aim to retain the refinement phase in the Leiden algorithm due to its beneficial properties, such as preventing the formation of poorly connected or even internally disconnected communities \cite{com-traag19}. To this end, unlike DF Louvain \cite{sahu2024shared}, we do not stop the algorithm after the first pass, but rather, run the algorithm until convergence occurs in a subsequent local-moving phase, where all vertices are processed, not just the affected ones. We refer to this method as \textbf{Full Refine}. Figure \ref{fig:subrefine-stages--02} shows an example of the communities returned with the \textit{Full Refine} method.

\subsection{Subset Refine method}
\label{sec:subset-refine-method}

Nevertheless, for small batch updates, only a few communities are typically affected, and only those need refinement. To identify these communities, we track vertices that migrate between communities during the local-moving phase and mark both their source and target communities as changed (i.e., needing refinement) because their sub-community structures may have been altered. However, the community membership IDs assigned to vertices during the local-moving phase can sometimes be inconsistent. For instance, a community labeled as $C = i$ might not contain a vertex with ID $i$ if that vertex has migrated to another community. This inconsistency can create problems during the refinement phase, where each vertex in the communities being refined must initially belong to its own community. To understand the issue, consider an example illustrated in Figure \ref{fig:subrefine-issue--1}. Here, say we selectively refine community $B$, which includes vertex $i$, but leave $C = i$ unchanged. In such a scenario, the vertex $i$ may remain as an isolated sub-community disconnected from the rest of $C = i$. Worse, as shown in Figure \ref{fig:subrefine-issue--2}, vertices $j$ and $k$ could merge with vertex $i$ to form a sub-community $D = i$, resulting in two unconnected communities, $C = i$ and $D = i$, sharing the same ID. This issue arises only when refining a subset of communities and does not occur if all communities are refined. To address this, we must assign each community $C = i$ a new ID $i'$, where $i'$ corresponds to one of its constituent vertices. Additionally, we must update the total community edge weights and changed community flags to reflect the new community IDs, as these are initially tied to the old IDs. The psuedocode for this is given in Algorithm \ref{alg:leiden-subset-renumber}, with its detailed explanation in Section \ref{sec:leiden-subset-renumber}.

\input{src/fig-subrefine-issue}

Next, we consider the impact of batch updates on the communities requiring refinement. Edge deletions within a community can cause it to split. However, if a community is isolated, the local-moving phase cannot find better community assignments for any of the vertices belonging to the community. In the extreme case, edge deletions within a community can also cause a community to be internally disconnected, if vertices within the community do not change their community assignments. This scenario can occur even if the community is not isolated. In order to address both concerns, refinement is the only way for such communities to split. Since community migrations do not occur, these communities would not be marked for refinement. Accordingly, for edge deletions in the batch update belonging to the same community, we mark the community as changed --- ensuring it is refined after the local-moving phase. Similarly, edge insertions affecting different parts of a community might also lead to a split, and must therefore be refined. Accordingly, for edge insertions in the batch update belonging to the same community, we mark the community as changed. Figure \ref{fig:community-split} shows an example of how these corner cases might occur with isolated communities, which may cause the community to split. We refer to this method of selective refinement of communities obtained from the local-moving phase of the algorithm, considering both vertex migration and the batch update, as \textbf{Subset Refine}. Figure \ref{fig:subrefine-stages--03} shows an example of the communities returned with the \textit{Subset Refine} method. Note that only a subset of the communities are processed into sub-communities.

Note that selective refinement is applied only in the first pass of the Leiden algorithm, while all communities are refined in subsequent passes. This is because untouched communities are already aggregated into super-vertices during the first pass. In the later passes, tracking communities across the super-vertex hierarchy would be costly, and the time saved by selective refinement in these smaller graphs wouldn't justify the overhead.

\input{src/fig-community-split}

\subsection{Optimized Aggregation method}
\label{sec:optimized-aggregation-method}

However, the above selective refinement method hardly improves the performance of the DF Leiden, particularly for small batch updates. For these smaller updates, the aggregation phase remains a major bottleneck. This is mainly due to the selective refinement of communities, which results in a large difference in the sizes of communities to be aggregated --- the refined communities tend to be small, while the unrefined communities tend to be large. This creates a heavy workload for threads aggregating these unrefined communities into super-vertices. To ensure better load balancing, heavily loaded threads should be assigned minimal additional communities. Dynamic work scheduling can help, with each thread being assigned a smaller range of community IDs, or chunks. However, chunk sizes that are too small can introduce significant scheduling overhead. Through experimentation with chunk sizes from $1$ to $2048$ --- on large real-world graphs (given in Table \ref{tab:dataset-large}), with uniformly random batch updates consisting of $80\%$ edge insertions and $20\%$ edge deletions (see Section \ref{sec:batch-generation}), on batch sizes of $10^{-7}|E|$, $10^{-5}|E|$, and $10^{-7}|E|$, where $|E|$ is the number of edges in the original graph --- we find that a chunk size of $32$ provides the best overall performance for ND, DS, and DF Leiden; as shown in Figure \ref{fig:aggregation-adjust-chunksize}. We refer to this method as \textbf{Optimized Aggregation}. Figure \ref{fig:subrefine-optimize} illustrates the\ignore{mean} runtime of DF Leiden based on\ignore{the three methods mentioned,} \textit{Full Refine}, \textit{Subset Refine}, and \textit{Optimized Aggregation} methods, on large graphs with batch updates of size $10^{-7}|E|$ to $0.1|E|$.

\input{src/fig-aggregation-adjust-chunksize}
\input{src/fig-subrefine-optimize}

The pseudocode for \textit{Optimized Aggregation} based ND, DS, and DF Leiden, which we from here on refer to simply as ND, DS, and DF Leiden, respectively, is presented in Algorithms \ref{alg:naive}, \ref{alg:delta}, and \ref{alg:frontier}, with detailed explanations in Sections \ref{sec:our-naive}, \ref{sec:our-delta}, \ref{sec:our-frontier}, respectively. In the first pass of Leiden algorithm, we process the vertices identified as affected by the DS and DF approaches, initializing the community membership of each vertex based on the membership obtained from the previous snapshot of the graph. In the following passes, all super-vertices are marked as affected and processed according to the Leiden algorithm \cite{sahu2024shared}. Similar to DF Louvain \cite{sahu2024dflouvain}, we use the weighted degrees of vertices and the total edge weights of communities as auxiliary information \cite{sahu2024dflouvain}\ignore{, as illustrated in Figure \ref{fig:about-auxiliary}}. Note that guarantees of the original Leiden algorithm \cite{com-traag19} extend to our methods as selective refinement only prunes untouched communities, while processing remaining communities identically to original Leiden.

\subsection{Our Parallel Dynamic Frontier (DF) Leiden}
\label{sec:our-frontier}

Algorithm \ref{alg:frontier} presents the pseudocode for our Parallel Dynamic Frontier (DF) Leiden. It takes as input the updated graph snapshot $G^t$, edge deletions $\Delta^{t-}$ and insertions $\Delta^{t+}$ from the batch update, the previous community assignments $C^{t-1}$ for each vertex, the previous weighted degrees $K^{t-1}$ of vertices, and the previous total edge weights $\Sigma^{t-1}$ of communities. It outputs the updated community memberships $C^t$ for vertices, the updated weighted degrees $K^t$ of vertices, and the updated total edge weights $\Sigma^t$ of communities.

In the algorithm, we initially identify a set of affected vertices whose communities may directly change due to batch updates by marking them in the flag vector $\delta V$. This is achieved by marking the endpoints of edge deletions $\Delta^{t-}$ that lie within the same community (lines \ref{alg:frontier--loopdel-begin}-\ref{alg:frontier--loopdel-end}), and by marking the endpoints of edge insertions $\Delta^{t+}$ that lie in disjoint communities (lines \ref{alg:frontier--loopins-begin}-\ref{alg:frontier--loopins-end}). Subsequently, three lambda functions are defined for the Leiden algorithm: \texttt{isAffected()} (lines \ref{alg:frontier--isaff-begin}-\ref{alg:frontier--isaff-end}), \texttt{inAffectedRange()} (lines \ref{alg:frontier--isaffrng-begin}-\ref{alg:frontier--isaffrng-end}), and \texttt{onChange()} (lines \ref{alg:frontier--onchg-begin}-\ref{alg:frontier--onchg-end}). These functions indicate that a set of vertices are initially marked as affected, that all vertices in the graph can be incrementally marked as affected, and that the neighbors of a vertex are marked as affected if it changes its community membership, respectively. It is important to note that the set of affected vertices will expand automatically due to vertex pruning optimization used in our Parallel Leiden algorithm (Algorithm \ref{alg:leiden}). Thus, \texttt{onChange()} reflects what the DF approach would do in the absence of vertex pruning. Additionally, unlike existing approaches, we utilize $K^{t-1}$ and $\Sigma^{t-1}$, alongside the batch updates $\Delta^{t-}$ and $\Delta^{t+}$, to efficiently compute $K^t$ and $\Sigma^t$ required for the local-moving phase of the Leiden algorithm (line \ref{alg:frontier--auxiliary}). These lambda functions and the total vertex/edge weights are then employed to execute the Leiden algorithm and obtain the updated community assignments $C^t$ (line \ref{alg:frontier--leiden}). Finally, we return $C^t$, alongside $K^t$ and $\Sigma^t$\ignore{, serving as the updated auxiliary information} (line \ref{alg:frontier--return}).

\input{src/alg-frontier}

\subsection{Our Dynamic-supporting Parallel Leiden}
\label{sec:our-leiden}

The main step of our Dynamic-supporting Parallel Leiden is outlined in Algorithm \ref{alg:leiden}. In contrast to our implementation of Static Leiden \cite{sahu2023gveleiden}, this algorithm, in addition to the current graph snapshot $G^t$, accepts the prior community membership of each vertex $C^{t-1}$, the revised weighted degree of each vertex $K^t$, the updated total edge weight of each community $\Sigma^t$, and a set of lambda functions $F$ determining whether a vertex is influenced or can be progressively identified as influenced (within the affected range). It yields the updated community memberships of vertices $C^t$.

\input{src/alg-leiden}

In the algorithm, we start by marking the affected vertices as unprocessed (lines \ref{alg:leiden--mark-begin}-\ref{alg:leiden--mark-end}). Next, during the initialization phase, several parameters are set: the community membership of each vertex $C$ in $G^t$, the total edge weight of each vertex $K'$, the total edge weight of each community $\Sigma'$, and the community membership $C'$ of each vertex in the current (super-vertex) graph $G'$. Additionally, an initial set of changed communities $\Delta C'$ is determined based on the batch update as described in Section \ref{sec:approach} (lines \ref{alg:leiden--init-begin}-\ref{alg:leiden--init-end}). Following initialization, a series of passes is carried out, limited by $MAX\_PASSES$, with each pass comprising local-moving, refinement, and aggregation phases (lines \ref{alg:leiden--passes-begin}-\ref{alg:leiden--passes-end}). In each pass, the Leiden algorithm’s local-moving phase (Algorithm \ref{alg:leidenlm}) is executed to optimize community assignments (line \ref{alg:leiden--local-move}). However, this phase may result in poorly connected or internally disconnected communities. Therefore, the communities must be refined. For selective refinement --- which enhances the algorithm's efficiency for small batch updates --- the community IDs are renumbered so that each community ID corresponds to one of its constituent vertices. Additionally, the total edge weights of communities and the flags for changed communities are updated accordingly (see Algorithm \ref{alg:leidensr}), and changed communities are broken up, isolating their constituent vertices into separate communities. The total weights of these communities are adjusted as needed. The refinement phase, executed by \texttt{leidenRefine()}, optimizes the community assignments for each vertex within its community boundaries $C'_B$ obtained from the local-moving phase (line \ref{alg:leiden--refine}). If the local-moving phase converges after a single iteration, indicating global convergence, the passes are halted (line \ref{alg:leiden--globally-converged}). If convergence is not achieved, the aggregation phase is performed. This involves renumbering the communities, updating top-level community memberships $C$ using dendrogram lookup, performing the aggregation process (Algorithm \ref{alg:leidenag}), and updating the weighted degrees of vertices $K'$ and total edge weights of communities $\Sigma'$ in the super-vertex graph. Following this, to prepare for the next pass of the algorithm, we mark all vertices in this graph as unprocessed, initialize community memberships based on the refinement phase, mark all communities as changed for the next pass, and adjust the convergence threshold $\tau$ by scaling it (line \ref{alg:leiden--threshold-scaling}). The next pass then begins (line \ref{alg:leiden--passes-begin}). After all passes, the top-level community memberships $C$ of each vertex in $G^t$ are updated one final time via dendrogram lookup before being returned (lines \ref{alg:leiden--lookup-last}-\ref{alg:leiden--return}).

We now discuss about two support functions \textsc{changedCommunities()} and \textsc{breakChangedCommunities()}. The function \textsc{changedCommunities()} takes as input the batch update, consisting of edge deletions $\Delta^{t-}$ and insertions $\Delta^{t+}$, and marks a community as changed if both the endpoints of edge deletions and insertions belong to the same community. However, this is only done for ND, DS, and DF Leiden --- all communities are considered to haved changed with Static Leiden. Finally, the function \textsc{breakChangedCommunities()} splits changed communities, as indicated with the changed communities flag $\Delta C'$, into vertices belonging to isolated communities.

\subsubsection{Renumbering communities by ID of a vertex within}
\label{sec:leiden-subset-renumber}

We now describe Algorithm \ref{alg:leidensr}, which outlines a method for renumbering communities based on their internal vertices. The goal is to ensure that each community is identified by the ID of one of its member vertices. The inputs to the algorithm include the current graph $G'$, the vertex community assignments $C'$, the total edge weights of communities $\Sigma'$, and the changed communities flags $\Delta C'$.

The algorithm begins by initializing several essential data structures ($C''$, $\Sigma''$, $\Delta C''$, and $C'_v$) to empty or default values (line \ref{alg:leidensr--init}). These structures will store updated community memberships, edge weights, change flags, and help select a representative vertex for each community. Next, for each vertex $i$, the current community ID $c'$ is retrieved from $C'[i]$. If no representative vertex has been assigned yet for community $c'$ (i.e., $C'_v[c']$ is empty), vertex $i$ is designated as the representative for community $c'$. Following this, the second parallel loop processes all communities $c' \in \Gamma'$, the set of communities in the original graph. For each community $c'$, the representative vertex $c''$ is retrieved from the previous step, and the total edge weight $\Sigma'[c']$ and the change flag $\Delta C'[c']$ are reorganized into the corresponding updated versions $\Sigma''[c'']$ and $\Delta C''[c'']$. Finally, the third parallel loop updates the community membership of each vertex based on the representative vertex chosen for its community. For each vertex $i$, the algorithm finds the current community $C'[i]$, retrieves the representative vertex for that community, and reassigns vertex $i$ to the new community ID corresponding to its representative. Once the community memberships are updated, the algorithm performs an in-place update of the original structures: the community memberships $C'$, total edge weights $\Sigma'$, and change flags $\Delta C'$ are replaced with the updated versions $C''$, $\Sigma''$, and $\Delta C''$.

\input{src/alg-leidensr}

\subsection{Implementation details}
\label{sec:implementation-details}

We use an asynchronous version of Leiden, where threads independently process different graph parts, enabling faster convergence but increasing variability in the final result. Each thread has its own hashtable to track delta-modularity during the local-moving and refinement phases and the total edge weight between super-vertices in the aggregation phase. Optimizations include OpenMP's dynamic loop schedule, limiting iterations to $20$ per pass, a tolerance drop rate of $10$ starting at $0.01$, vertex pruning, using parallel prefix sums and preallocated Compressed Sparse Row (CSR) data structures for super-vertex graphs and community vertices, and using fast, collision-free per-thread hashtables\ignore{for all algorithm phases} \cite{sahu2024fast}.

For simplicity, we do not skip the aggregation phase of Leiden algorithm, even if a small number of communities are being merged together, unlike the original implementation of Static Leiden \cite{sahu2024fast}. This allows us to maintain a high quality of obtained communities while incurring only a minor increase in runtime across Static, ND, DS, and DF Leiden. Additionally, we employ the refine-based variation of Leiden, where the community labels of super-vertices are determined by the refinement phase rather than the local-moving phase (i.e., move-based)\ignore{\cite{sahu2024fast}}, as the latter would not permit community splits in the scenarios outlined in Section \ref{sec:subset-refine-method}.

\subsection{Time and Space complexity}

The time complexity of ND, DS, and DF Leiden is the same as Static Leiden, i.e., $O(L|E^t|)$, where $L$ is the total number of iterations performed. However, the cost of local-moving and refinement phases in the first pass is reduced, and depends both on the size and nature of the batch update. The space complexity of our algorithms is also the same as Static Leiden, i.e., $O(T|V^t| + |E^t|)$, where $T$ represents the number of threads used ($T|V^t|$ is the space used by per-thread collision-free hashtables \cite{sahu2024fast}).

%% file: src/fig-subrefine-stages.tex
\begin{figure*}[hbtp]
  \centering
  \subfigure[No continued passes]{
    \label{fig:subrefine-stages--01}
    \includegraphics[width=0.31\linewidth]{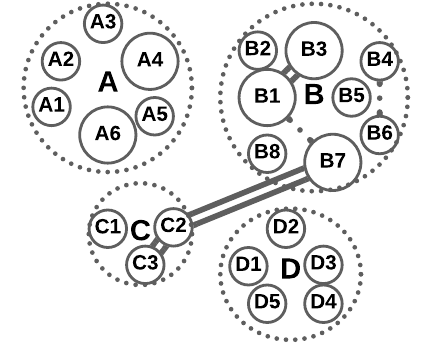}
  }
  \subfigure[Full Refine]{
    \label{fig:subrefine-stages--02}
    \includegraphics[width=0.31\linewidth]{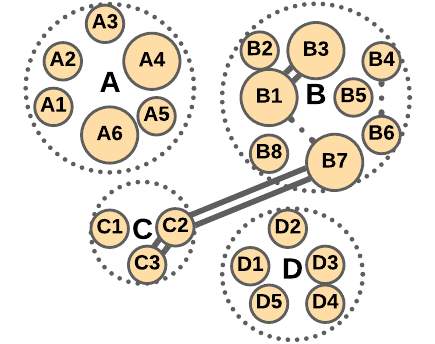}
  }
  \subfigure[Subset Refine]{
    \label{fig:subrefine-stages--03}
    \includegraphics[width=0.31\linewidth]{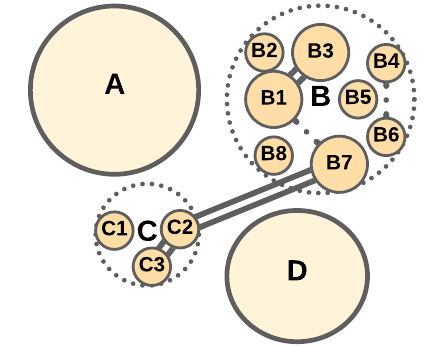}
  } \\[-1ex]
  \caption{Comparison of \textit{No continued passes}, \textit{Full Refine}, and \textit{Subset Refine} methods. Here, circles represent communities (or subcommunities post refinement), dotted circles denote old parent communities during the local-moving phase\ignore{community bounds}, dotted lines indicate edge deletions, double lines signify edge insertions, and a brown fill indicates mandated further processing. Pre-existing edges are not shown\ignore{for simplicity}. With \textit{No continued passes}, all communities are refined after the local-moving phase; however, it may converge prematurely with small batch updates, resulting in suboptimal community structures. The \textit{Full Refine} method processes all refined communities after they are aggregated into super-vertices. In contrast, with the \textit{Subset Refine} method, we selectively refine only a subset of communities based on the batch update, leaving the remaining communities unchanged.}
  \label{fig:subrefine-stages}
\end{figure*}

%% file: src/fig-subrefine-issue.tex
\begin{figure}[hbtp]
  \centering
  \subfigure[Disconnected isolated vertex]{
    \label{fig:subrefine-issue--1}
    \includegraphics[width=0.45\linewidth]{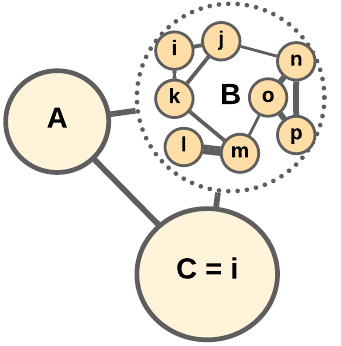}
  }
  \subfigure[Disconnected community formation]{
    \label{fig:subrefine-issue--2}
    \includegraphics[width=0.45\linewidth]{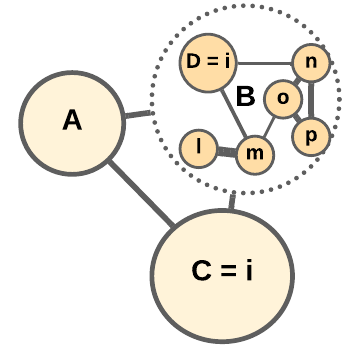}
  } \\[-1ex]
  \caption{Illustration of issues arising during the refinement phase when only a subset of communities is refined. Here, circles represent communities (or subcommunities after refinement), dotted circles indicate old parent communities (from the local-moving phase), and lines show both inter- and intra-community edges. Upon refinement of community $B$, subfigure (a) shows that vertex $i$ is isolated, but disconnected from community $C = i$, while subfigure (b) shows that further refinement forms sub-community $D = i$, which is disconnected from community $C = i$.}
  \label{fig:subrefine-issue}
\end{figure}

%% file: src/fig-community-split.tex
\begin{figure}[hbtp]
  \centering
  \subfigure{
    \label{fig:community-split--all}
    \includegraphics[width=0.80\linewidth]{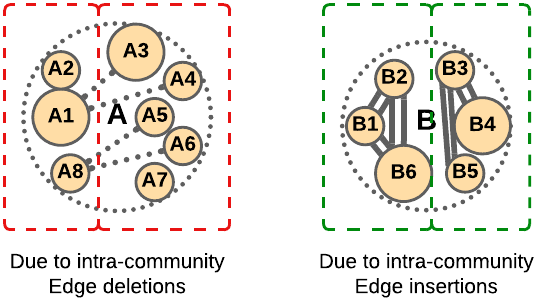}
  } \\[-1ex]
  \caption{Demonstration of how decreasing or increasing edge density within a community can cause it to split. Here, circles show refined subcommunities, while dotted circles represent the original parent communities from the local-moving phase. Dotted lines indicate edge deletions, double lines represent edge insertions, and brown-filled areas mark regions needing further processing. Red and green boundaries highlight possible split points due to batch updates.}
  \label{fig:community-split}
\end{figure}

%% file: src/fig-aggregation-adjust-chunksize.tex
\begin{figure}[!htp]
  \centering
  \subfigure[Relative runtimes on uniformly random batch updates of size $10^{-7}|E|$]{
    \label{fig:aggregation-adjust-chunksize--batch7}
    \includegraphics[width=0.98\linewidth]{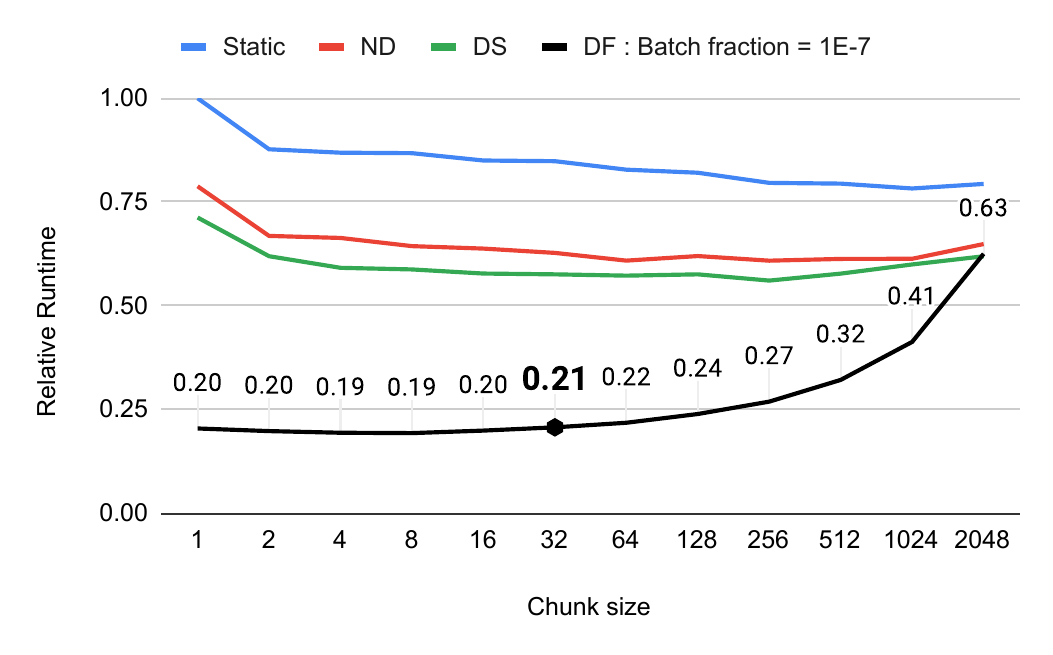}
  }
  \subfigure[Relative runtimes on uniformly random batch updates of size $10^{-5}|E|$]{
    \label{fig:aggregation-adjust-chunksize--batch5}
    \includegraphics[width=0.98\linewidth]{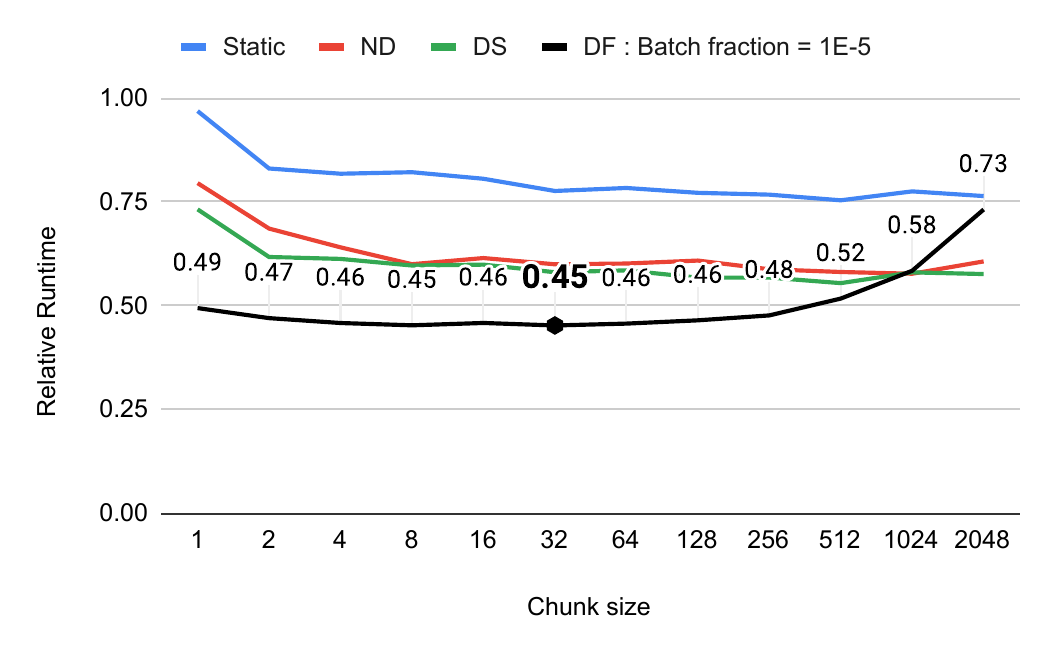}
  }
  \subfigure[Relative runtimes on uniformly random batch updates of size $10^{-3}|E|$]{
    \label{fig:aggregation-adjust-chunksize--batch3}
    \includegraphics[width=0.98\linewidth]{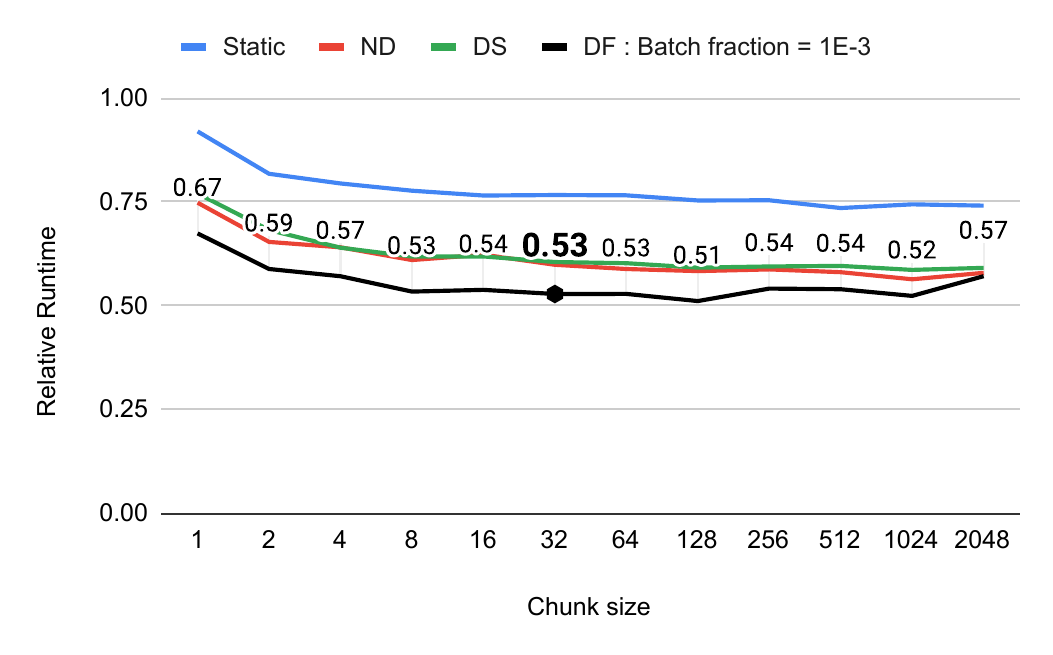}
  } \\[-1ex]
  \caption{Relative Runtime of \textit{Static}, \textit{Naive-dynamic (ND)}, \textit{Delta-screening (DS)}, and \textit{Dynamic Frontier (DF)} Leiden, with varying dynamic schedule chunk size (OpenMP), for aggregation phase of the Leiden algorithm. These tests were conducted on large graphs, with batch updates randomly generated at sizes of $10^{-7}|E|$, $10^{-5}|E|$, and $10^{-3}|E|$. The results suggest that a chunk size of $32$ is optimal (highlighted). In this figure, relative runtimes are normalized to maximum runtime, specifically that of Static Leiden with a chunk size of $1$ for dynamic scheduling during the aggregation phase.}
  \label{fig:aggregation-adjust-chunksize}
\end{figure}

%% file: src/fig-subrefine-optimize.tex
\begin{figure}[!hbt]
  \centering
  \subfigure{
    \label{fig:subrefine-optimize--8020}
    \includegraphics[width=0.98\linewidth]{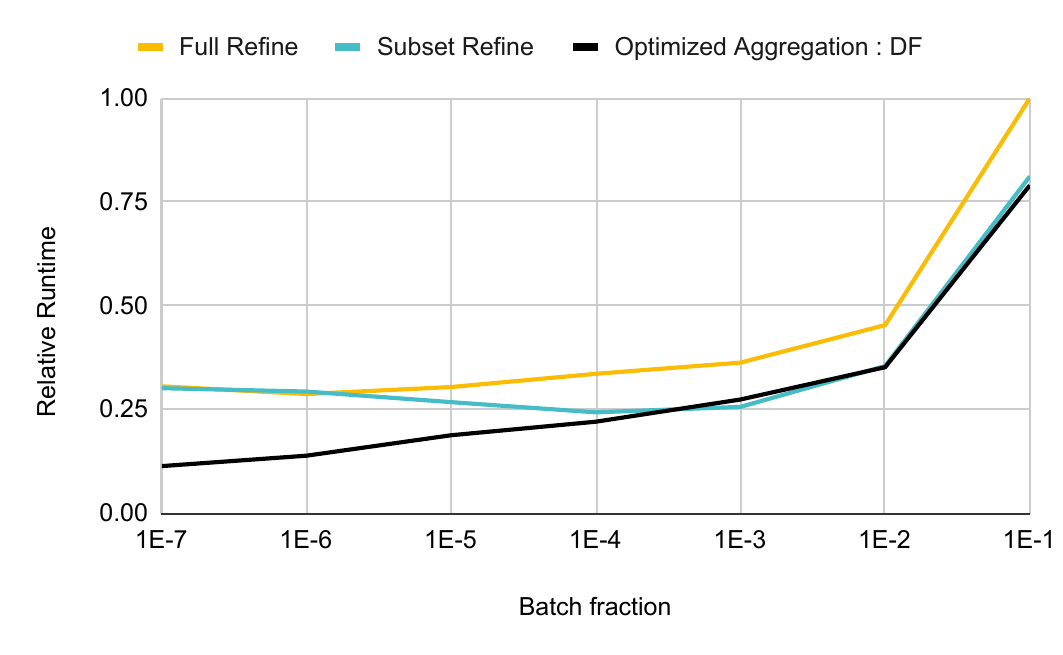}
  } \\[-2ex]
  \caption{Relative Runtime of \textit{Dynamic Frontier (DF)} approach applied to the Leiden algorithm, incorporating three successive optimizations: \textit{Full Refine}, \textit{Subset Refine}, and \textit{Optimized Aggregation}. These are tested on large graphs with randomly generated batch updates of size $10^{-7}|E|$ to $0.1|E|$, consisting of $80\%$ edge insertions and $20\%$ deletions.}
  \label{fig:subrefine-optimize}
\end{figure}

%% file: src/alg-frontier.tex
\begin{algorithm}[hbtp]
\caption{Our Parallel \textit{Dynamic Frontier (DF)} Leiden \cite{sahu2024dflouvain}.}
\label{alg:frontier}
\begin{algorithmic}[1]
\Require{$G^t(V^t, E^t)$: Current/updated input graph}
\Require{$\Delta^{t-}, \Delta^{t+}$: Edge deletions and insertions (batch update)}
\Require{$C^{t-1}, C^t$: Previous, current community of each vertex}
\Require{$K^{t-1}, K^t$: Previous, current weighted-degree of vertices}
\Require{$\Sigma^{t-1}, \Sigma^t$: Previous, current total edge weight of communities}
\Ensure{$\delta V$: Flag vector indicating if each vertex is affected}
\Ensure{$isAffected(i)$: Is vertex $i$ is marked as affected?}
\Ensure{$inAffectedRange(i)$: Can $i$ be incrementally marked?}
\Ensure{$onChange(i)$: What happens if $i$ changes its community?}
\Ensure{$F$: Lambda functions passed to parallel Leiden (Alg. \ref{alg:leiden})}

\Statex

\Function{dynamicFrontier}{$G^t, \Delta^{t-}, \Delta^{t+}, C^{t-1}, K^{t-1}, \Sigma^{t-1}$}
  \State $\rhd$ Mark initial affected vertices
  \ForAll{$(i, j) \in \Delta^{t-}$ \textbf{in parallel}} \label{alg:frontier--loopdel-begin}
    \If{$C^{t-1}[i] = C^{t-1}[j]$} $\delta V[i] \gets 1$
    \EndIf
  \EndFor \label{alg:frontier--loopdel-end}
  \ForAll{$(i, j, w) \in \Delta^{t+}$ \textbf{in parallel}} \label{alg:frontier--loopins-begin}
    \If{$C^{t-1}[i] \neq C^{t-1}[j]$} $\delta V[i] \gets 1$
    \EndIf
  \EndFor \label{alg:frontier--loopins-end}
  \Function{isAffected}{$i$} \label{alg:frontier--isaff-begin}
    \Return{$\delta V[i]$}
  \EndFunction \label{alg:frontier--isaff-end}
  \Function{inAffectedRange}{$i$} \label{alg:frontier--isaffrng-begin}
    \Return{$1$}
  \EndFunction \label{alg:frontier--isaffrng-end}
  \Function{onChange}{$i$} \label{alg:frontier--onchg-begin}
    \ForAll{$j \in G^t.neighbors(i)$} $\delta V[j] \gets 1$
    \EndFor
  \EndFunction \label{alg:frontier--onchg-end}
  \State $F \gets \{isAffected, inAffectedRange, onChange\}$ \label{alg:frontier--lambdas}
  \State $\rhd$ Use $K^{t-1}$, $\Sigma^{t-1}$ as auxiliary information (Alg. \ref{alg:update})
  \State $\{K^t, \Sigma^t\} \gets updateWeights(G^t, \Delta^{t-}, \Delta^{t+}, C^{t-1}, K^{t-1}, \Sigma^{t-1})$\label{alg:frontier--auxiliary}
  \State $\rhd$ Obtain updated communities (Alg. \ref{alg:leiden})
  \State $C^t \gets leiden(G^t, C^{t-1}, K^t, \Sigma^t, F)$ \label{alg:frontier--leiden}
  \Return{$\{C^t, K^t, \Sigma^t\}$} \label{alg:frontier--return}
\EndFunction
\end{algorithmic}
\end{algorithm}

%% file: src/alg-leiden.tex
\begin{algorithm}[hbtp]
\caption{Our Dynamic-supporting Parallel Leiden \cite{sahu2024fast}.}
\label{alg:leiden}
\begin{algorithmic}[1]
\Require{$G^t(V^t, E^t)$: Current input graph}
\Require{$\Delta^{t-}, \Delta^{t+}$: Edge deletions and insertions (batch update)}
\Require{$C^{t-1}$: Previous community of each vertex}
\Require{$K^t$: Current weighted-degree of each vertex}
\Require{$\Sigma^t$: Current total edge weight of each community}
\Require{$F$: Lambda functions passed to parallel Leiden}
\Ensure{$G'(V', E')$: Current/super-vertex graph}
\Ensure{$C, C'$: Current community of each vertex in $G^t$, $G'$}
\Ensure{$K, K'$: Current weighted-degree of each vertex in $G^t$, $G'$}
\Ensure{$\Sigma, \Sigma'$: Current total edge weight of each community in $G^t$, $G'$}
\Ensure{$C'_B$: Community bound of each vertex}
\Ensure{$\Delta C'$: Changed communities flag}
\Ensure{$\tau$: Iteration tolerance}

\Statex

\Function{leiden}{$G^t, C^{t-1}, K^t, \Sigma^t, F$} \label{alg:leiden--begin}
  \State $\rhd$ Mark affected vertices as unprocessed
  \ForAll{$i \in V^t$} \label{alg:leiden--mark-begin}
    \If{$F.isAffected(i)$} Mark $i$ as unprocessed
    \EndIf
  \EndFor \label{alg:leiden--mark-end}
  \State $\rhd$ Initialization phase
  \State Vertex membership: $C \gets [0 .. |V^t|)$ \label{alg:leiden--init-begin}
  \State $G' \gets G^t$ \textbf{;} $C' \gets C^{t-1}$ \textbf{;} $K' \gets K^t$ \textbf{;} $\Sigma' \gets \Sigma^t$
  \State $\Delta C' \gets changedCommunities(G^t, C^{t-1}, \Delta^{t-}, \Delta^{t+})$ \label{alg:leiden--init-end}
  \State $\rhd$ Local-moving and aggregation phases
  \ForAll{$l_p \in [0 .. \text{MAX\_PASSES})$} \label{alg:leiden--passes-begin}
    \State $l_i \gets leidenMove(G', C', K', \Sigma', \Delta C', F)$ \Comment{Alg. \ref{alg:leidenlm}} \label{alg:leiden--local-move}
    \State $leidenSubsetRenumber(G', C', \Sigma', \Delta C')$ \Comment{Alg. \ref{alg:leidensr}} \label{alg:leiden-subset-renumber}
    \State $C'_B \gets C'$ \Comment{Community bounds for refinement phase}
    \State $breakChangedCommunities(G', C', K', \Sigma', \Delta C')$ \label{alg:leiden--reset-again}
    \State $leidenRefine(G', C'_B, C', K', \Sigma', \tau)$ \Comment{Alg. \ref{alg:leidenre}} \label{alg:leiden--refine}
    \If{\textbf{not} first pass \textbf{and} $l_i \le 1$} \textbf{break} \Comment{Done?} \label{alg:leiden--globally-converged}
    \EndIf
    \State $C' \gets$ Renumber communities in $C'$ \label{alg:leiden--renumber}
    \State $C \gets$ Lookup dendrogram using $C$ to $C'$ \label{alg:leiden--lookup}
    \State $G' \gets leidenAggregate(G', C')$ \Comment{Alg. \ref{alg:leidenag}} \label{alg:leiden--aggregate}
    \State $\Sigma' \gets K' \gets$ Weight of each vertex in $G'$ \label{alg:leiden--reset-weights}
    \State Mark all vertices in $G'$ as unprocessed \label{alg:leiden--reset-affected}
    \State $C' \gets [0 .. |V'|)$ \Comment{Use refine-based membership} \label{alg:leiden--useparent}
    \State $\Delta C' \gets \{1\ \forall\ V'\}$ \Comment{Refine all communities next pass} \label{alg:leiden--all-communities-changed}
    \State $\tau \gets \tau / \text{TOLERANCE\_DROP}$ \Comment{Threshold scaling} \label{alg:leiden--threshold-scaling}
  \EndFor \label{alg:leiden--passes-end}
  \State $C \gets$ Lookup dendrogram using $C$ to $C'$ \label{alg:leiden--lookup-last}
  \Return{$C$} \label{alg:leiden--return}
\EndFunction \label{alg:leiden--end}

\Statex

\Function{changedCommunities}{$G^t, C^{t-1}, \Delta^{t-}, \Delta^{t+}$}
  \State $\Delta C' \gets \{\}$
  \If{is dynamic alg.}
    \ForAll{$(i, j) \in \Delta^{t-} \cup \Delta^{t+}$ \textbf{in parallel}}
      \If{$C^{t-1}[i] = C^{t-1}[j]$} $\Delta C'[i] \gets 1$
      \EndIf
    \EndFor
  \Else\ $\Delta C' \gets \{1\ \forall\ V^t\}$
  \EndIf
  \Return{$\Delta C'$}
\EndFunction

\Statex

\Function{breakChangedCommunities}{$G', C', K', \Sigma', \Delta C'$}
  \ForAll{$i \in V'$ \textbf{in parallel}}
    \If{$\Delta C'[C'[i]] = 0$} \textbf{continue}
    \EndIf
    \State $C'[i] \gets i$ \textbf{;} $\Sigma'[i] \gets K'[i]$
  \EndFor
\EndFunction
\end{algorithmic}
\end{algorithm}


%% file: src/alg-leidensr.tex
\begin{algorithm}[hbtp]
\caption{Renumber communities by ID of a vertex within.}
\label{alg:leidensr}
\begin{algorithmic}[1]
\Require{$G'(V', E')$: Input/super-vertex graph}
\Require{$C', C''$: Current, updated community membership of vertices}
\Require{$\Sigma', \Sigma''$: Current, updated total edge weight of each community}
\Require{$\Delta C', \Delta C''$: Current, updated changed communities flag}
\Ensure{$\Gamma'$: Set of communities in $C'$}

\Statex

\Function{leidenSubsetRenumber}{$G', C', \Sigma', \Delta C'$}
  \State $C'' \gets \Sigma'' \gets \Delta C'' \gets C'_v \gets \{\}$ \label{alg:leidensr--init}
  \State $\rhd$ Obtain any vertex from each community
  \ForAll{$i \in V'$ \textbf{in parallel}}
    \State $c' \gets C'[i]$
    \If{$C'_v[c] = \text{EMPTY}$} $C'_v[c] \gets i$
    \EndIf
  \EndFor
  \State $\rhd$ Update community weights and changed status
  \ForAll{$c' \in \Gamma'$ \textbf{in parallel}}
    \State $c'' \gets C'_v[c']$
    \If{$c'' \neq \text{EMPTY}$}
      \State $\Sigma''[c''] \gets \Sigma'[c']$
      \State $\Delta C''[c''] \gets \Delta C'[c']$
    \EndIf
  \EndFor
  \State $\rhd$ Update community memberships
  \ForAll{$i \in V'$ \textbf{in parallel}}
    \State $C''[i] \gets C'_v[C'[i]]$
  \EndFor
  \State $\rhd$ Update in-place
  \State $C' \gets C''$ \textbf{;} $\Sigma' \gets \Sigma''$ \textbf{;} $\Delta C' \gets \Delta C''$
\EndFunction
\end{algorithmic}
\end{algorithm}

%% file: 05-evaluation.tex
\subsection{Experimental setup}
\label{sec:setup}

\subsubsection{System}
\label{sec:system}

\ignore{In our experiments, }We employ a server equipped with an x86-based 64-core AMD EPYC-7742 processor. It operates at a clock frequency of $2.25$ GHz and is paired with $512$ GB of DDR4 system memory. Each core features a $4$ MB L1 cache, a $32$ MB L2 cache, and a shared L3 cache of $256$ MB. The server runs on Ubuntu 20.04.

\subsubsection{Configuration}
\label{sec:configuration}

\input{src/tab-dataset-large}
\input{src/tab-dataset}

We utilize 32-bit unsigned integers for vertex IDs and 32-bit floating-point for edge weights. For computation involving floating-point aggregation, like total edge weights and modularity, we switch to 64-bit floating-point. Affected vertices and changed communities are represented by 8-bit integer vectors. During the local-moving, refinement, and aggregation phases, we employ OpenMP's dynamic scheduling with a chunk size of $2048$ for load balancing. However, for the aggregation phase in Naive-dynamic (ND), Delta-screening (DS), and Dynamic Frontier (DF) Leiden, a chunk size of $32$ is used. We utilize state-of-the-art parallel Leiden \cite{sahu2024fast}, incorporating our ND, DS, and DF Leiden atop it. For real-world dynamic graphs (discussed in Section \ref{sec:performance-comparison-temporal}), which aggregate slowly, we disable aggregation tolerance - setting it to $1$. This ensures all desirable community aggregations occur, yielding higher modularity scores compared to the default of $0.8$. For consistency, we apply this to static Leiden as well - to avoid comparing apples-to-oranges. We run all\ignore{parallel} implementations on $64$ threads and compile using GCC 9.4 with OpenMP 5.0.

\subsubsection{Dataset}
\label{sec:dataset}

We conduct experiments on $12$ large real-world graphs with random batch updates, listed in Table \ref{tab:dataset-large}, obtained from the SuiteSparse Matrix Collection. These graphs have between $3.07$ million and $214$ million vertices, and $25.4$ million to $3.80$ billion edges. For experiments involving real-world dynamic graphs, we utilize five temporal networks sourced from the Stanford Large Network Dataset Collection \cite{snapnets}, detailed in Table \ref{tab:dataset}. Here, the number of vertices span from $24.8$ thousand to $2.60$ million, temporal edges from $507$ thousand to $63.4$ million, and static edges from $240$ thousand to $36.2$ million. However, we would like to note that existing temporal graphs in the SNAP repository \cite{snapnets} are mostly small\ignore{(with one exception)}, limiting their utility for studying our proposed parallel algorithms. With each graph, we ensure that all edges are undirected and weighted, with a default weight of $1$. Since most publicly available real-world weighted graphs are small in size, we do not use them here --- although our parallel algorithms can handle weighted graphs without any modification.

We also do not use SNAP datasets with ground-truth communities, as they are non-disjoint, while our work focuses on disjoint communities. Nevertheless, we wish to note that the goal of community detection is not necessarily to match the ground truth, as this may reflect a different aspect of network structure or be unrelated to the actual structure, and focusing solely on this correlation risks overlooking other meaningful community structures \cite{peel2017ground}.

\subsubsection{Batch generation}
\label{sec:batch-generation}

We take each base graph from Table \ref{tab:dataset-large} and generate random batch updates \cite{com-zarayeneh21}, comprising an $80\% : 20\%$ mix of edge insertions and deletions to emulate realistic batch updates, with each edge having a weight of $1$. Batch updates are generated with unit-weighted edges for simplicity, without affecting results, as the algorithms handle arbitrary edge weights.\ignore{In the batch updates, $??\%$ edge deletions are intra-community, while $??\%$ insertions are cross-community, reflecting strong connectivity of identified communities.} Since the random batch updates are processed separately, the graph does not converge toward a random structure. This enables testing on diverse graph classes. All updates are undirected; for every edge insertion $(i, j, w)$, we also include $(j, i, w)$. To prepare the set of edges for insertion, we select vertex pairs with equal probability. For edge deletions, we uniformly delete existing edges. To simplify, no new vertices are added or removed from the graph. The batch size is measured as a fraction of the edges in the original graph, ranging from $10^{-7}$ to $0.1$ (i.e., $10^{-7}|E|$ to $0.1|E|$). For a billion-edge graph, this translates to a batch size ranging from $100$ to $100$ million edges. We employ five distinct random batch updates for each batch size and report the average across these runs. Note that dynamic graph algorithms are helpful for small batch updates in interactive applications. For large batches, it is usually more efficient to run the static algorithm.

\subsubsection{Measurement}
\label{sec:measurement}

We evaluate the runtime of each method on the entire updated graph, encompassing all stages: the local-moving phase, refinement phase, aggregation phase, initial and incremental marking of affected vertices, convergence detection, and any necessary intermediate steps.\ignore{We exclude memory allocation and deallocation times.} To mitigate the impact of noise, we follow the standard practice of running each experiment multiple times. We assume that the total edge weight of the graphs is known and can be tracked with each batch update. As our baseline, we use the most efficient multicore implementation of Static Leiden \cite{sahu2024fast}, which has been shown to outperform even GPU-based implementations. Given that modularity maximization is NP-hard, all existing polynomial-time algorithms are heuristic; thus, we assess the optimality of our dynamic algorithms by comparing their convergence to the modularity score achieved by the static algorithm. Finally, none of the analyzed algorithms --- Static, ND, DS, and DF Leiden --- produce communities that are internally disconnected, so we omit this from our figures.

\input{src/fig-8020-runtime}
\input{src/fig-8020-modularity}

\subsection{Performance Comparison}
\label{sec:performance-comparison}

We now evaluate the performance of our parallel implementations of ND, DS, and DF Leiden against Static Leiden \cite{sahu2024fast}, on large graphs with randomly generated batch updates. Static Leiden is rerun from scratch after each update, whereas ND, DS, and DF Leiden begin with prior community assignments. Additionally, DS and DF Leiden identify affected vertices, processing only these during local-moving phase. Our methods are applicable to ND, DS, and DF Leiden. Original Leiden \cite{com-traag19} is sequential; thus we compare against the fastest parallel Leiden \cite{sahu2024fast}. As detailed in Section \ref{sec:batch-generation}, these updates range in size from $10^{-7}|E|$ to $0.1|E|$, with $80\%$ edge insertions and $20\%$ edge deletions. Each batch update includes reverse edges to keep the graph undirected. Figure \ref{fig:8020-runtime--all} presents the execution time of each algorithm across individual graphs, while Figure \ref{fig:8020-runtime--mean} depicts overall runtimes using geometric mean for consistent scaling across differing graph sizes. In addition, Figure \ref{fig:8020-modularity--all} shows the modularity of the algorithms for individual graphs in the dataset, and Figure \ref{fig:8020-runtime--mean} presents the overall modularity of each of the algorithms, calculated using arithmetic mean.

From Figure \ref{fig:8020-runtime--mean}, we observe that ND, DS, and DF Leiden achieve mean speedups of $1.37\times$, $1.47\times$, and $1.98\times$, respectively, when compared to Static Leiden. This speedup is higher on smaller batch updates, with ND, DS, and DF Leiden being on average $1.46\times$, $1.74\times$, and $3.72\times$, respectively, when compared to Static Leiden, on batch updates of size $10^{-7}|E|$. Figure \ref{fig:8020-runtime--all} shows that ND, DS, and DF Leiden offer moderate speedups over Static Leiden on web graphs and social networks --- while on road networks (with low average degree), ND Leiden marginally outperforms Static Leiden, and DS and DF Leiden perform significantly better. This is due to DS and DF Leiden processing significantly fewer affected vertices/communities on road networks. On protein k-mer graphs, DS Leiden offers only a slight improvement. However, on small batch updates, DF Leiden outperforms Static Leiden by a large margin.

In terms of modularity, as Figures \ref{fig:8020-modularity--mean} and \ref{fig:8020-modularity--all} show, ND, DS, and DF Leiden achieve communities with approximately the same modularity as Static Leiden. However, on social networks, ND, DS, and DF Leiden moderately outperform Static Leiden. This is likely due to social networks not having a strong enough community structure, and thus requiring more iterations to converge to a better community assignment. On such graphs, Static Leiden fails to attain better modularity due to it limiting the number of iterations per pass performed, in favor of faster runtimes. However, since ND, DS, and DF Leiden do not start from scratch, but leverage prior community assignments, they are able to converge faster, and with higher modularity than Static Leiden (Figure \ref{fig:8020-modularity}) --- particularly on social networks, where poor community structures make clustering more challenging. Thus, on average, the modularity of communities from Static Leiden is slightly lower. However, we notice that modularity of communities identified by our algorithms do not differ by more than $0.002$ from Static Leiden, on average.

Also note in Figure \ref{fig:8020-runtime} that the runtime of Static Leiden increases with larger batch updates. This effect is primarily due to two factors: \textbf{(1)} Larger batches increase graph size, as $80\%$ of updates are edge insertions; and \textbf{(2)} Random edge updates disrupt original community structure, requiring more iterations to converge. This\ignore{disruption} also accounts for the observed decline in modularity of the algorithms with larger batch sizes, as seen in Figure \ref{fig:8020-modularity}.

Let us now discuss why ND, DS, and DF Leiden exhibit only modest speedups over Static Leiden. Unlike Louvain algorithm, we must always run the refinement phase of Leiden algorithm to avoid badly connected and internally disconnected communities. Nonetheless, the refinement phase splits the communities obtained/updated from the local-moving phase into several smaller sub-communities. Stopping the passes early\ignore{, as done with DF Louvain,} leads to low modularity scores for ND/DS/DF Leiden because sufficiently large high-quality clusters have not yet formed\ignore{due to the refinement phase}. Further, ND/DS/DF Leiden can only reduce the runtime of the local-moving phase of the first pass of the Leiden algorithm. However, only about $37\%$ of the runtime in Static Leiden is spent in the local-moving phase of the first pass. These factors constrain the speedup potential of ND, DS, and DF Leiden over Static Leiden. Regardless, on large graphs with random batch updates, DF Leiden appears to be the dynamic community detection method of choice for tracking evolving communities. Performance comparison on real-world dynamic graphs is given in Section \ref{sec:performance-comparison-temporal}.

\subsection{\ignore{Analysis of }Affected vertices and Changed communities}

We now study the fraction of vertices marked as affected, and the communities marked as changed (to keeping track of the subset of communities to be refined), by DS and DF Leiden on instances from Table \ref{tab:dataset-large}, with batch updates ranging from $10^-7|E|$ to $0.1|E|$. This tracking occurs only in the first pass of the Leiden algorithm, as detailed in Sections \ref{sec:subset-refine-method} and \ref{sec:optimized-aggregation-method}, and is shown in Figure \ref{fig:8020-affected}.

\input{src/fig-8020-affected}
\input{src/fig-8020-scaling}

We notice from Figure \ref{fig:8020-affected} that DF Leiden marks fewer affected vertices and changed communities than DS Leiden, but their runtime difference is smaller, as many vertices marked by DS Leiden do not change their community labels, and converge quickly. However, as expected, performance of the algorithms decline with more affected vertices, as Figures \ref{fig:8020-runtime} and \ref{fig:8020-affected} show. Further, while a high number of changed communities increases refinement costs in the first pass, the cost of subsequent passes\ignore{is more or less constant, and} only depends on the graph's nature.

\subsection{Scalability}

Finally, we study the strong-scaling behavior of ND, DS, and DF Leiden, and compare it with Static Leiden. For this, we fix the batch size at $10^{-3} |E|$, vary the\ignore{number of} threads\ignore{in use} from $1$ to $64$, and measure the speedup of each algorithm with respect to its sequential version.

As shown in Figure \ref{fig:8020-scaling}, ND, DS, and DF Leiden obtain a speedup of $10.2\times$, $9.9\times$, and $9.0\times$ respectively at $32$ threads (with respect to sequential\ignore{execution}); with their speedup increasing at a mean rate of $1.59\times$, $1.58\times$, and $1.55\times$, respectively, for every doubling of threads. At $64$ threads, NUMA affects the performance of our algorithms (NPS 4). In addition, increasing the number of threads makes the behavior of the algorithm less asynchronous, i.e., similar to the Jacobi iterative method, further contributing to the performance drop.

%% file: src/tab-dataset-large.tex
\begin{table}[hbtp]
  \centering
  \caption{List of $12$ graphs retrieved from the SuiteSparse Matrix Collection \cite{suite19} (with directed graphs indicated by $*$). Here, $|V|$ denotes the number of vertices, $|E|$ denotes the number of edges (after making the graph undirected by adding reverse edges), and $|\Gamma|$ denotes the number of communities obtained with \textit{Static Leiden} algorithm \cite{sahu2024fast}.}
  \label{tab:dataset-large}
  \begin{tabular}{|c||c|c|c|}
    \toprule
    \textbf{Graph} &
    \textbf{\textbf{$|V|$}} &
    \textbf{\textbf{$|E|$}} &
    \textbf{\textbf{$|\Gamma|$}} \\
    \midrule
    \multicolumn{4}{|c|}{\textbf{Web Graphs (LAW)}} \\ \hline
    indochina-2004$^*$ & 7.41M & 341M & 2.68K \\ \hline
    arabic-2005$^*$ & 22.7M & 1.21B & 2.92K \\ \hline
    uk-2005$^*$ & 39.5M & 1.73B & 18.2K \\ \hline
    webbase-2001$^*$ & 118M & 1.89B & 2.94M \\ \hline
    it-2004$^*$ & 41.3M & 2.19B & 4.05K \\ \hline
    sk-2005$^*$ & 50.6M & 3.80B & 2.67K \\ \hline
    \multicolumn{4}{|c|}{\textbf{Social Networks (SNAP)}} \\ \hline
    com-LiveJournal & 4.00M & 69.4M & 3.09K \\ \hline
    com-Orkut & 3.07M & 234M & 36 \\ \hline
    \multicolumn{4}{|c|}{\textbf{Road Networks (DIMACS10)}} \\ \hline
    asia\_osm & 12.0M & 25.4M & 2.70K \\ \hline
    europe\_osm & 50.9M & 108M & 6.13K \\ \hline
    \multicolumn{4}{|c|}{\textbf{Protein k-mer Graphs (GenBank)}} \\ \hline
    kmer\_A2a & 171M & 361M & 21.1K \\ \hline
    kmer\_V1r & 214M & 465M & 10.5K \\ \hline
  \bottomrule
  \end{tabular}
\end{table}

%% file: src/tab-dataset.tex
\begin{table}[hbtp]
  \centering
  \caption{List of $5$ real-world dynamic graphs obtained from the Stanford Large Network Dataset Collection \cite{snapnets}. Here, $|V|$ denotes the vertex count, $|E_T|$ denotes the total number of temporal edges (including duplicates), and $|E|$ denotes the number of static edges (excluding duplicates).}
  \label{tab:dataset}
  \begin{tabular}{|c||c|c|c|c|}
    \toprule
    \textbf{Graph} &
    \textbf{\textbf{$|V|$}} &
    \textbf{\textbf{$|E_T|$}} &
    \textbf{\textbf{$|E|$}} \\
    \midrule
    sx-mathoverflow & 24.8K & 507K & 240K \\ \hline
    sx-askubuntu & 159K & 964K & 597K \\ \hline
    sx-superuser & 194K & 1.44M & 925K \\ \hline
    wiki-talk-temporal & 1.14M & 7.83M & 3.31M \\ \hline
    sx-stackoverflow & 2.60M & 63.4M & 36.2M \\ \hline
  \bottomrule
  \end{tabular}
\end{table}

%% file: src/fig-8020-runtime.tex
\begin{figure*}[hbtp]
  \centering
  \subfigure[Overall result]{
    \label{fig:8020-runtime--mean}
    \includegraphics[width=0.38\linewidth]{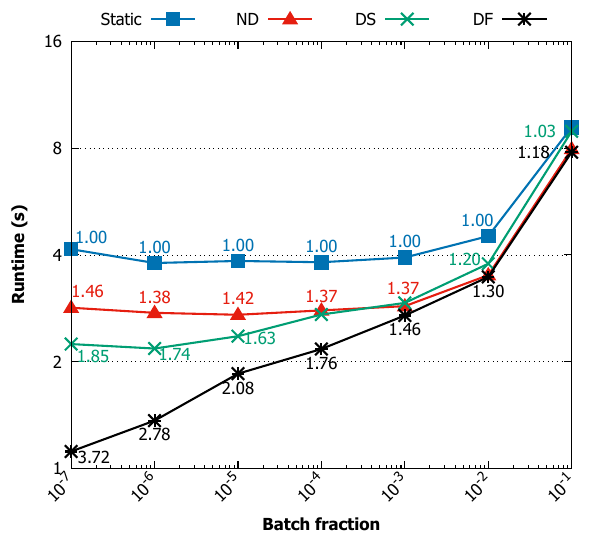}
  }
  \subfigure[Results on each graph]{
    \label{fig:8020-runtime--all}
    \includegraphics[width=0.58\linewidth]{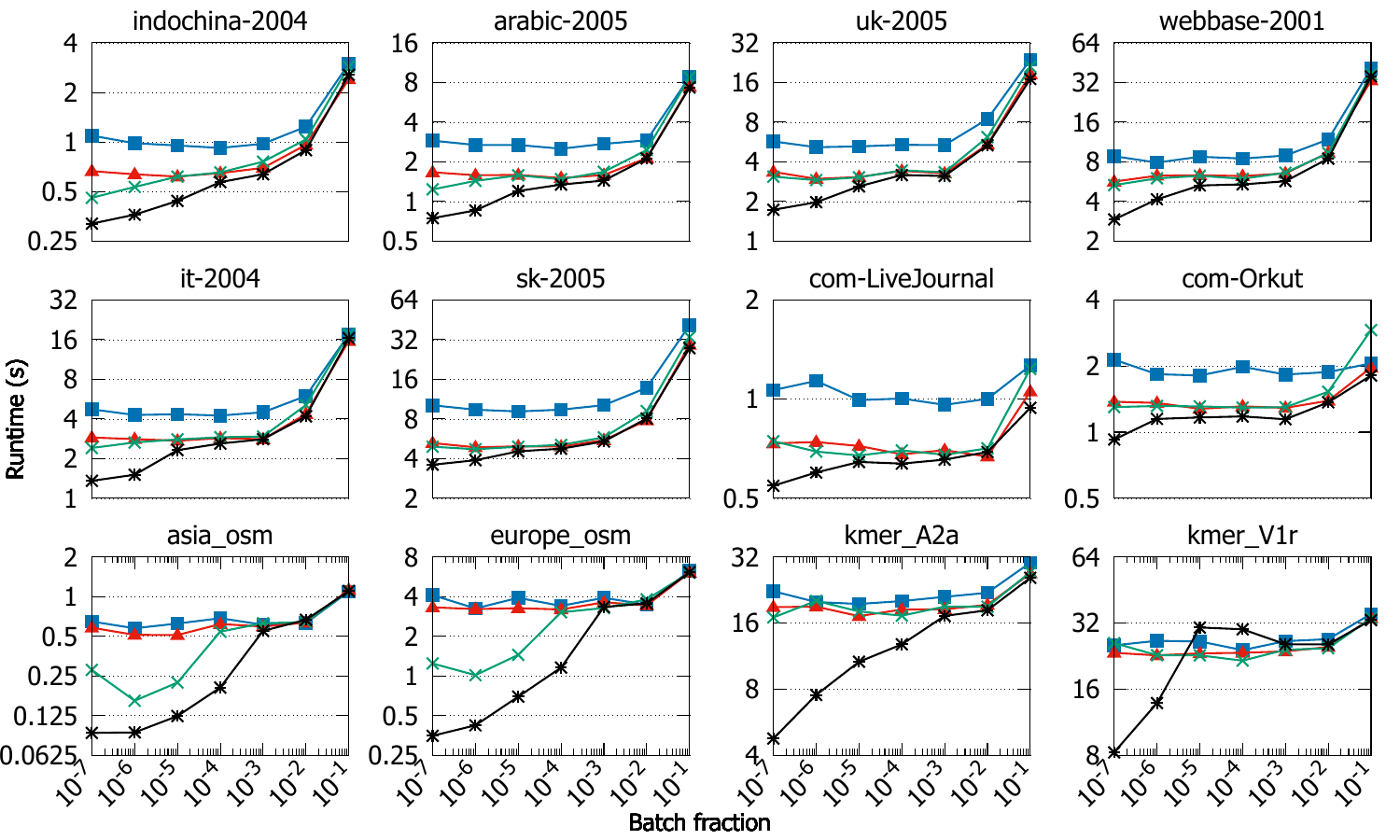}
  } \\[-2ex]
  \caption{Runtime (logarithmic scale) of our multicore implementation of \textit{Naive-dynamic (ND)}, \textit{Delta-screening (DS)}, and \textit{Dynamic Frontier (DF) Leiden}, compared to \textit{Static Leiden} \cite{sahu2024fast}, on large\ignore{(static)} graphs with randomly generated batch updates. The size of these batch updates ranges from $10^{-7}|E|$ to $0.1|E|$ in multiples of $10$, with the updates comprising $80\%$ edge insertions and $20\%$ edge deletions to simulate realistic dynamic graph changes. The right subfigure shows the runtime of each algorithm for individual graphs in the dataset, while the left subfigure displays overall runtimes using the geometric mean for consistent scaling across graphs.\ignore{Furthermore,} The speedup of each algorithm compared to Static Leiden is indicated on the respective lines.}
  \label{fig:8020-runtime}
\end{figure*}

%% file: src/fig-8020-modularity.tex
\begin{figure*}[hbtp]
  \centering
  \subfigure[Overall result]{
    \label{fig:8020-modularity--mean}
    \includegraphics[width=0.38\linewidth]{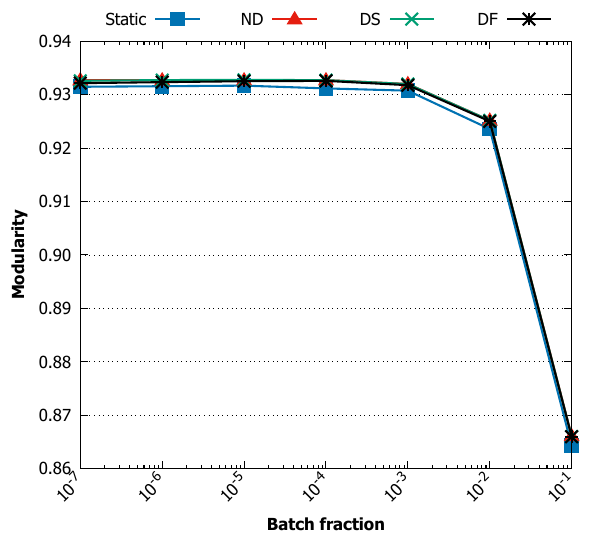}
  }
  \subfigure[Results on each graph]{
    \label{fig:8020-modularity--all}
    \includegraphics[width=0.58\linewidth]{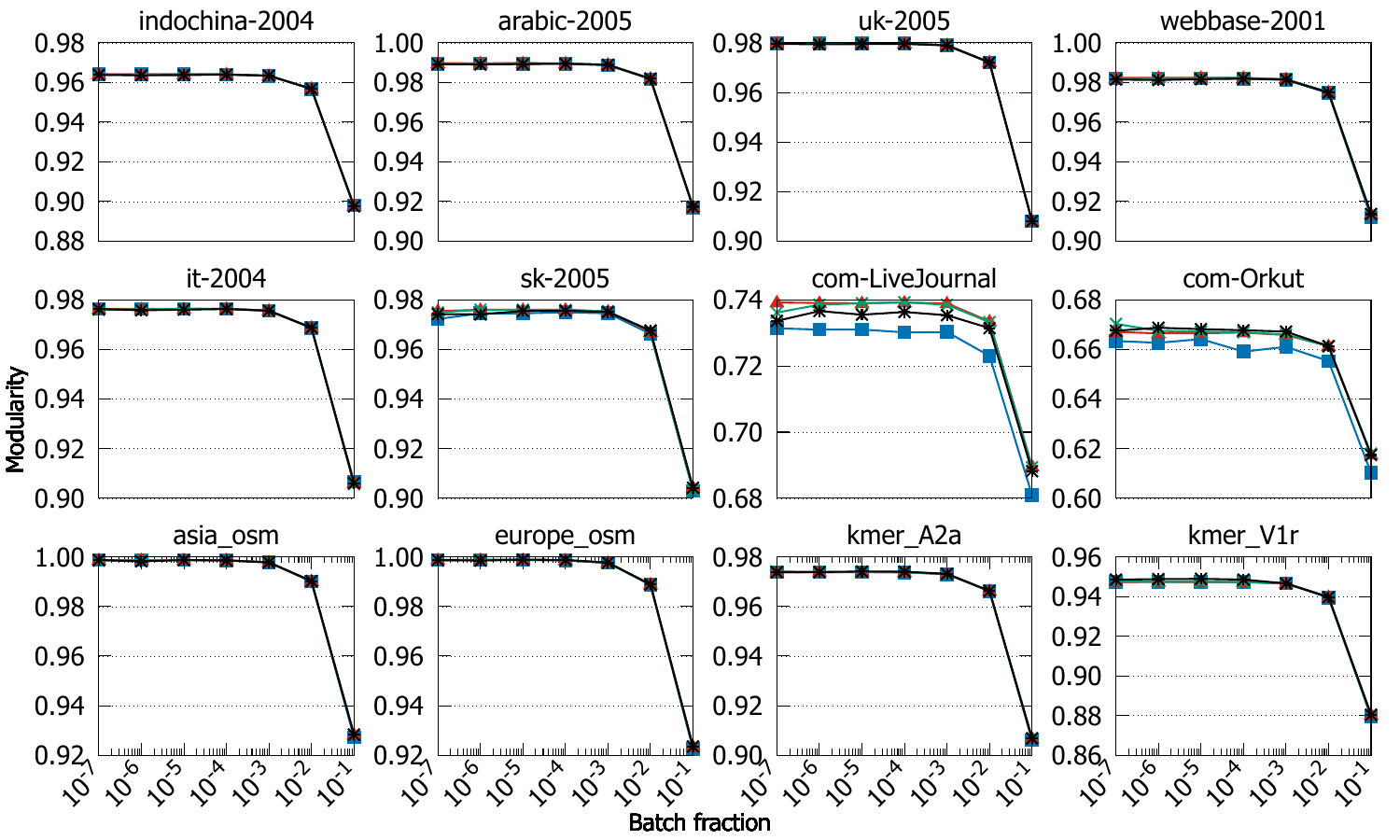}
  } \\[-2ex]
  \caption{Modularity comparison of our multicore implementation of \textit{Naive-dynamic (ND)}, \textit{Delta-screening (DS)}, and \textit{Dynamic Frontier (DF) Leiden}, compared to \textit{Static Leiden} \cite{sahu2024fast}, on large\ignore{(static)} graphs with randomly generated batch updates. These batch updates vary in size from $10^{-7}|E|$ to $0.1|E|$ in powers of 10, consisting of $80\%$ edge insertions and $20\%$ edge deletions to mimic realistic dynamic graph updates. Here, the right subfigure shows the modularity for each algorithm for individual graphs in the dataset, while the left subfigure displays overall modularity obtained using arithmetic mean.}
  \label{fig:8020-modularity}
\end{figure*}

%% file: src/fig-8020-affected.tex
\begin{figure}[hbtp]
  \centering
  \subfigure{
    \label{fig:8020-affected--all}
    \includegraphics[width=0.98\linewidth]{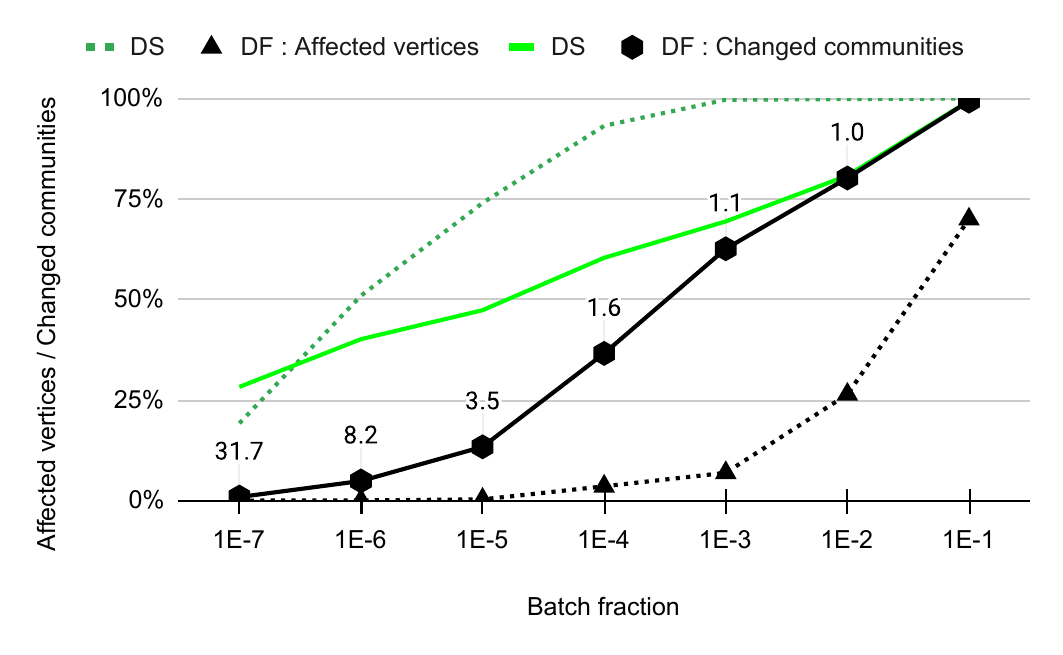}
  } \\[-4ex]
  \caption{Fraction of vertices marked as affected (dotted lines), and communities marked as changed (solid lines) with \textit{Delta-screening (DS)} and \textit{Dynamic Frontier (DF) Leiden} on graphs in Table \ref{tab:dataset-large}. The labels indicate the ratio of communities marked as changed by DS Leiden to that of DF Leiden.}
  \label{fig:8020-affected}
\end{figure}

%% file: src/fig-8020-scaling.tex
\begin{figure}[hbtp]
  \centering
  \subfigure{
    \label{fig:8020-scaling--am}
    \includegraphics[width=0.98\linewidth]{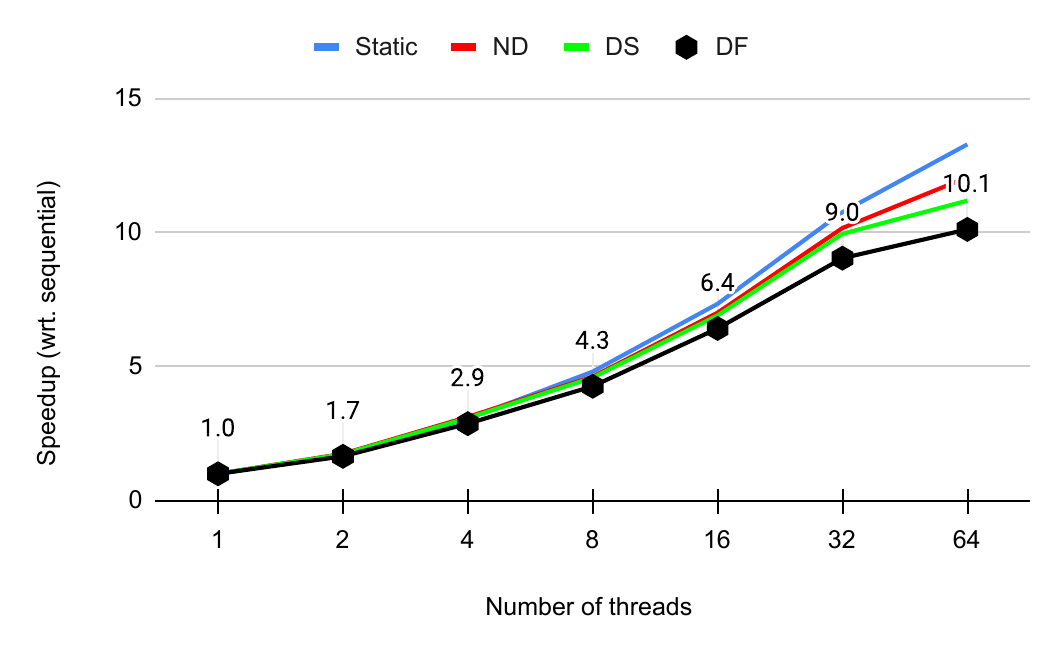}
  } \\[-4ex]
  \caption{Strong scalability of ND, DS, and DF Leiden, compared to Static Leiden, on batch updates of size $10^{-3} |E|$. The number of threads is doubled from $1$ to $64$ (logarithmic scale).}
  \label{fig:8020-scaling}
\end{figure}

%% file: 06-conclusion.tex
In this paper, we show how to adapt three approaches for community detection in dynamic graphs --- Naive-dynamic (ND), Delta-screening (DS), and Dynamic Frontier (DF) --- to a fast multicore implementation \cite{sahu2024fast} of the Leiden algorithm \cite{com-traag19}. This enables users to track community changes over time while reducing the computational cost of updating communities as the graph evolves. Our experiments on a server equipped with a 64-core AMD EPYC-7742 CPU show that ND, DS, and DF Leiden achieve moderate speedups of $1.37\times$, $1.47\times$, and $1.98\times$, respectively, on large graphs with randomized batch updates, compared to Static Leiden. These small gains are primarily because we need to run the algorithm for several passes, similar to Static Leiden, in order to obtain high quality top-level communities, and the reduction in runtime occurs only during the local-moving and refinement phases of the first pass of the algorithm. For every doubling of threads, our algorithms scale by $1.6\times$. We hope these\ignore{initial} results encourage further investigation of dynamic Leiden algorithm\ignore{in the context of evolving graphs}.

%% file: aa-appendix.tex
\subsection{Our Parallel Naive-dynamic (ND) Leiden}
\label{sec:our-naive}

Algorithm \ref{alg:naive} details our multicore implementation of Naive dynamic (ND) Leiden. Here, vertices are assigned to communities based on the previous snapshot of the graph, with all vertices being processed regardless of edge deletions and insertions in the batch update. The algorithm accepts the current/updated graph snapshot $G^t$, edge deletions $\Delta^{t-}$ and insertions $\Delta^{t+}$ in the batch update, the previous community membership $C^{t-1}$ for each vertex, the weighted degree of each vertex $K^{t-1}$, and the total edge weight of each community $\Sigma^{t-1}$. The output includes the updated community memberships $C^t$, the updated weighted degrees $K^t$, and the updated total edge weights of communities $\Sigma^t$.

In the algorithm, we begin by defining two lambda functions for the Leiden algorithm: \texttt{isAffected()} (lines \ref{alg:naive--isaff-begin} to \ref{alg:naive--isaff-end}) and \texttt{inAffected} \texttt{Range()} (lines \ref{alg:naive--isaffrng-begin} to \ref{alg:naive--isaffrng-end}). These functions indicate that all vertices in the graph $G^t$ should be marked as affected and that these vertices can be incrementally marked as affected, respectively. Unlike previous approaches, we then use $K^{t-1}$ and $\Sigma^{t-1}$, along with the batch updates $\Delta^{t-}$ and $\Delta^{t+}$, to quickly compute $K^t$ and $\Sigma^t$, which are required for the local-moving phase of the Leiden algorithm (line \ref{alg:naive--auxliliary}). The lambda functions, along with the total vertex and edge weights, are then employed to run the Leiden algorithm and obtain\ignore{the} updated community assignments $C^t$ (line \ref{alg:naive--leiden}). Finally, $C^t$ is returned, along with $K^t$ and $\Sigma^t$ as\ignore{the} updated auxiliary information (line \ref{alg:naive--return}).

\input{src/alg-naive}
\input{src/alg-delta}

\subsection{Our Parallel Delta-screening (DS) Leiden}
\label{sec:our-delta}

Algorithm \ref{alg:delta} presents the pseudocode for our multicore implementation of Delta-screening (DS) Leiden. It employs modularity-based scoring to identify an approximate region of the graph where vertices are likely to change their community membership \cite{com-zarayeneh21}. It takes as input the current/updated graph snapshot $G^t$, edge deletions $\Delta^{t-}$ and insertions $\Delta^{t+}$ in the batch update, the previous community memberships of vertices $C^{t-1}$, the weighted degrees of vertices $K^{t-1}$, and the total edge weights of communities $\Sigma^{t-1}$. The algorithm outputs the updated community memberships $C^t$, weighted degrees $K^t$, and total edge weights of communities $\Sigma^t$. Prior to processing, the batch update --- which includes edge deletions $(i, j, w) \in \Delta^{t-}$ and insertions $(i, j, w) \in \Delta^{t+}$ --- are sorted separately by the source vertex ID $i$.

In the algorithm, we start by initializing a hashtable $H$ that maps communities to their associated weights, and we set the affected flags $\delta V$, $\delta E$, and $\delta C$, which indicate whether a vertex, its neighbors, or its community is affected by the batch update (lines \ref{alg:delta--init}). We then parallelly process edge deletions $\Delta^{t-}$ and insertions $\Delta^{t+}$. For each deletion $(i, j, w) \in \Delta^{t-}$ where vertices $i$ and $j$ belong to the same community, we mark the source vertex $i$, its neighbors, and its community as affected (lines \ref{alg:delta--loopdel-begin}-\ref{alg:delta--loopdel-end}). For each unique source vertex $i$ in insertions $(i, j, w) \in \Delta^{t+}$ where $i$ and $j$ belong to different communities, we determine the community $c^*$ with the highest delta-modularity if $i$ moves to one of its neighboring communities, marking $i$, its neighbors, and the community $c^*$ as affected (lines \ref{alg:delta--loopins-begin}-\ref{alg:delta--loopins-end}). We disregard deletions between different communities and insertions within the same community. Using the affected neighbor $\delta E$ and community flags $\delta C$, we mark affected vertices in $\delta V$ (lines \ref{alg:delta--loopaff-begin}-\ref{alg:delta--loopaff-end}). Subsequently, similar to ND Leiden, we utilize $K^{t-1}$ and $\Sigma^{t-1}$, along with $\Delta^{t-}$ and $\Delta^{t+}$, to quickly derive $K^t$ and $\Sigma^t$ (line \ref{alg:naive--auxliliary}). We define the necessary lambda functions \texttt{isAffected()} (lines \ref{alg:delta--isaff-begin}-\ref{alg:delta--isaff-end}) and \texttt{inAffectedRange()} (lines \ref{alg:delta--isaffrng-begin}-\ref{alg:delta--isaffrng-end}), and execute the Leiden algorithm, resulting in updated community assignments $C^t$ (line \ref{alg:delta--leiden}). Finally, we return\ignore{the} updated community memberships $C^t$ along with $K^t$ and $\Sigma^t$ as\ignore{the} updated auxiliary information (line \ref{alg:delta--return}).

\subsection{Our Dynamic-supporting Parallel Leiden}
\label{sec:our-leiden-continued}

\subsubsection{Local-moving phase of our Parallel Leiden}

The pseudocode for the local-moving phase\ignore{of our Parallel Leiden} is presented in Algorithm \ref{alg:leidenlm}. It iteratively moves vertices among communities in order to maximize modularity. Here, the \texttt{leidenMove()} function operates on the current graph $G'$, community membership $C'$, total edge weight of each vertex $K'$, total edge weight of each community $\Sigma'$, changed communities flag vector $\Delta C'$, and a set of lambda functions as inputs, yielding the number of iterations performed $l_i$.

\input{src/alg-leidenlm}

Lines \ref{alg:leidenlm--iterations-begin}-\ref{alg:leidenlm--iterations-end} encapsulate the primary loop of the local-moving phase. In line \ref{alg:leidenlm--init-deltaq}, we initialize the total delta-modularity per iteration $\Delta Q$. Subsequently, in lines \ref{alg:leidenlm--loop-vertices-begin}-\ref{alg:leidenlm--loop-vertices-end}, we concurrently iterate over unprocessed vertices. For each vertex $i$, we perform vertex pruning by marking $i$ as processed (line \ref{alg:leidenlm--prune}). Next, we verify if $i$ falls within the affected range (i.e., it is permitted to be incrementally marked as affected), and if not, we proceed to the next vertex (line \ref{alg:leidenlm--affrng}). For each unskipped vertex $i$, we scan communities connected to $i$ (line \ref{alg:leidenlm--scan}), excluding itself, ascertain the optimal community $c*$ to move $i$ to (line \ref{alg:leidenlm--best-community-begin}), compute the delta-modularity of moving $i$ to $c*$ (line \ref{alg:leidenlm--best-community-end}), update the community membership of $i$ (lines \ref{alg:leidenlm--perform-move-begin}-\ref{alg:leidenlm--perform-move-end}), and mark its neighbors as unprocessed (line \ref{alg:leidenlm--remark}) if a superior community is identified. It is worth noting that this practice of marking neighbors of $i$ as unprocessed, which is part of the vertex pruning optimization, also aligns with algorithm of DF Leiden --- which marks its neighbors as affected, when a vertex changes its community. Thus, vertex pruning facilitates incremental expansion of the set of affected vertices without requiring any extra code. Further, for ND, DS, and DF Leiden, we mark the source $c$ and target communities $c^*$ of the migrating vertex $i$ as changed. With Static Leiden, all communities are always refined. In line \ref{alg:leidenlm--locally-converged}, we examine whether the local-moving phase has achieved convergence (locally); if so, the loop is terminated (or if $MAX\_ITERATIONS$ is reached). Finally, in line \ref{alg:leidenlm--return}, we return the number of iterations performed\ignore{by the local-moving phase} $l_i$.

\subsubsection{Refinement phase of our Parallel Leiden}

\input{src/alg-leidenre}

The pseudocode outlining the refinement phase of our Parallel Leiden is presented in Algorithm \ref{alg:leidenlm}. This phase closely resembles the local-moving phase but incorporates the community membership obtained for each vertex as a \textit{community bound}. In this phase, each vertex is required to select a community within its community bound to join, aiming to maximize modularity through iterative movements between communities, akin to the local-moving phase. At the onset of the refinement phase, the community membership of each vertex is reset so that each vertex initially forms its own community. The \texttt{leidenRefine()} function is employed, taking as input the current graph $G'$, the community bound of each vertex $C'_B$, the initial community membership $C'$ of each vertex, the total edge weight of each vertex $K'$, the initial total edge weight of each community $\Sigma'$, changed communities flag vector $\Delta C'$, and the current tolerance per iteration $\tau$, and returns the number of iterations executed $l_j$.

Lines \ref{alg:leidenre--loop-vertices-begin}-\ref{alg:leidenre--loop-vertices-end} embody the central aspect of the refinement phase. During this phase, we execute what is termed the constrained merge procedure \cite{com-traag19}. The essence of this procedure lies in enabling vertices, within each community boundary, to create sub-communities solely by permitting isolated vertices (i.e., vertices belonging to their own community) to alter their community affiliation. This process divides any internally-disconnected communities identified during the local-moving phase and prevents the emergence of new disconnected communities. Specifically, for every isolated vertex $i$ (line \ref{alg:leidenre--check-isolated}), we explore communities connected to $i$ within the \textit{same community boundary} - excluding itself (line \ref{alg:leidenre--scan}). The refinement phase is skipped for a vertex if its community has not been flagged as changed ($\Delta C'[c] = 0$), or if another vertex has joined its community, as indicated by the total community weight no longer matching the total edge weight of the vertex. Subsequently, we determine the optimal community $c*$ for relocating $i$ (line \ref{alg:leidenre--best-community-begin}), and assess the delta-modularity of transferring $i$ to $c*$ (line \ref{alg:leidenre--best-community-end}). If a superior community is identified, we attempt to update the community affiliation of $i$ provided it remains isolated (lines \ref{alg:leidenre--perform-move-begin}-\ref{alg:leidenre--perform-move-end}). Note that we do not migrate the vertex $i$ to community $c^*$ if the vertex representing $c^*$ has itself moved to another community.

\subsubsection{Aggregation phase of our Parallel Leiden}

The pseudocode for the aggregation phase is presented in Algorithm \ref{alg:leidenag}, wherein communities are merged into super-vertices. Specifically, the \texttt{leidenAggre} \texttt{gate()} function within this algorithm takes the current graph $G'$ and the community membership $C'$ as inputs and produces the super-vertex graph $G''$.

In the algorithm, the process begins by obtaining the offsets array for the community vertices in the CSR format, denoted as $G'_{C'}.offsets$, within lines \ref{alg:leidenag--coff-begin} to \ref{alg:leidenag--coff-end}. This starts with counting the number of vertices in each community using \texttt{countCommunityVertices()}, followed by performing an exclusive scan on the resulting array. Next, within lines \ref{alg:leidenag--comv-begin} to \ref{alg:leidenag--comv-end}, we concurrently traverse all vertices, atomically placing the vertices associated with each community into the community graph CSR $G'_{C'}$. Following this, the offsets array for the super-vertex graph CSR is computed by estimating the degree of each super-vertex within lines \ref{alg:leidenag--yoff-begin} to \ref{alg:leidenag--yoff-end}. This involves calculating the total degree of each community using \texttt{communityTotalDegree()}, followed by another exclusive scan. As a result, the super-vertex graph CSR is organized with intervals for the edges and weights array of each super-vertex. Then, in lines \ref{alg:leidenag--y-begin} to \ref{alg:leidenag--y-end}, we iterate over all communities $c \in [0, |\Gamma|)$ in parallel using dynamic loop scheduling, with a chunk size of $2048$ for both Static Leiden and a chunk size of $32$ for ND, DS, and DF Leiden. During this phase, all communities $d$ (and their respective edge weights $w$) connected to each vertex $i$ in community $c$ are included (via \texttt{scanCommunities()}, described in Algorithm \ref{alg:leidenlm}) in the per-thread hashtable $H_t$. Once $H_t$ contains all the connected communities and their weights, they are atomically added as edges to super-vertex $c$ in the super-vertex graph $G''$. Finally, in line \ref{alg:leidenag--return}, we return the super-vertex graph $G''$.

\input{src/alg-leidenag}

\subsection{Updating vertex/community weights}
\label{sec:our-update}

We will now elaborate on the parallel algorithm designed to compute the updated weighted degree of each vertex $K^t$ and the total edge weight of each community $\Sigma^t$. This algorithm operates based on the previous community memberships of vertices $C^{t-1}$, the weighted degrees of vertices $K^{t-1}$, the total edge weights of communities, and the batch update, which encompasses edge deletions $\Delta^{t-}$ and insertions $\Delta^{t+}$. The pseudocode for this algorithm is presented in Algorithm \ref{alg:update}.

In the algorithm, initialization of $K$ and $\Sigma$, representing the weighted degree of each vertex and the total edge weight of each community, respectively, occurs first (line \ref{alg:update--init}). Subsequently, utilizing multiple threads, we iterate over sets of edge deletions $\Delta^{t-}$ (lines \ref{alg:update--loopdel-begin}-\ref{alg:update--loopdel-end}) and edge insertions $\Delta^{t+}$ (lines \ref{alg:update--loopins-begin}-\ref{alg:update--loopins-end}). For each edge deletion $(i, j, w)$ in $\Delta^{t-}$, we ascertain the community $c$ of vertex $i$ based on the previous community assignment $C^{t-1}$ (line \ref{alg:update--delc}). If vertex $i$ belongs to the current thread's work-list, its weighted degree is decremented by $w$ (line \ref{alg:update--delk}), and if community $c$ belongs to the work-list, its total edge weight is also decremented by $w$ (line \ref{alg:update--delsigma}). Similarly, for each edge insertion $(i, j, w)$ in $\Delta^{t+}$, adjustments are made to the weighted degree of vertex $i$ and the total edge weight of its community. Finally, updated values of $K$ and $\Sigma$ for each vertex and community are returned for further processing (line \ref{alg:update--return}).

\input{src/alg-update}

\subsection{Performance Comparison on Real-world dynamic graphs}
\label{sec:performance-comparison-temporal}

We also evaluate the performance of our parallel implementations of Static, ND, DS, and DF Leiden on real-world dynamic graphs listed in Table \ref{tab:dataset}. These evaluations are performed on batch updates ranging from $10^{-5}|E_T|$ to $10^{-3}|E_T|$ in multiples of $10$. For each batch size, as described in Section \ref{sec:batch-generation}, we load $90\%$ of the graph, add reverse edges to ensure all edges are undirected, and then load $B$ edges (where $B$ is the batch size) consecutively in $100$ batch updates. Figure \ref{fig:temporal-summary--runtime-overall} shows the overall runtime of each approach across all graphs for each batch size, while Figure \ref{fig:temporal-summary--modularity-overall} depicts the overall modularity of the obtained communities. Additionally, Figures \ref{fig:temporal-summary--runtime-graph} and \ref{fig:temporal-summary--modularity-graph} present the mean runtime and modularity of the communities obtained with each approach on individual dynamic graphs in the dataset.

Figure \ref{fig:temporal-summary--runtime-overall} illustrates that ND Leiden is, on average, $1.09\times$ faster than Static Leiden for batch updates ranging from $10^{-5}|E_T|$ to $10^{-3}|E_T|$. In comparison, DS and DF Leiden exhibit average speedups of $1.20\times$ and $1.33\times$, respectively, over Static Leiden for the same batch updates. We now explain why ND, DS, and DF Leiden achieve only minor speedups over Static Leiden. Our experiments indicate that only about $20\%$ of the overall runtime of Static Leiden is spent in the local-moving phase of the first pass.\ignore{This is a significant decrease from the $37\%$ observed on large graphs with random batch updates.} Consequently, the speedup achieved by ND, DS, and DF Leiden relative to Static Leiden is limited. Furthermore, our observations suggest that while DF Leiden can reduce the time spent in the local-moving phase slightly more than ND Leiden, it incurs increased runtimes in the refinement and aggregation phases of the algorithm. This is likely because ND Leiden is able to optimize clusters more effectively than DF Leiden, which accelerates the refinement and aggregation phases for ND Leiden, albeit by a small margin. In addition, we observe that only about $20\%$ of the overall runtime of Static Leiden is spent in the local-moving phase of the first pass. This explains the lower speedup of DF Leiden compared to ND Leiden on larger batch updates of real-world dynamic graphs.\ignore{Therefore, we recommend ND Leiden for real-world dynamic graphs.}

\input{src/fig-temporal-summary}

%% file: src/alg-naive.tex
\begin{algorithm}[hbtp]
\caption{Our Parallel \textit{Naive-dynamic (ND)} Leiden \cite{sahu2024dflouvain}.}
\label{alg:naive}
\begin{algorithmic}[1]
\Require{$G^t(V^t, E^t)$: Current/updated input graph}
\Require{$\Delta^{t-}, \Delta^{t+}$: Edge deletions and insertions (batch update)}
\Require{$C^{t-1}, C^t$: Previous, current community of each vertex}
\Require{$K^{t-1}, K^t$: Previous, current weighted-degree of vertices}
\Require{$\Sigma^{t-1}, \Sigma^t$: Previous, current total edge weight of communities}
\Ensure{$isAffected(i)$: Is vertex $i$ is marked as affected?}
\Ensure{$inAffectedRange(i)$: Can $i$ be incrementally marked?}
\Ensure{$F$: Lambda functions passed to parallel Leiden (Alg. \ref{alg:leiden})}

\Statex

\Function{naiveDynamic}{$G^t, \Delta^{t-}, \Delta^{t+}, C^{t-1}, K^{t-1}, \Sigma^{t-1}$}
  \State $\rhd$ Mark affected vertices
  \Function{isAffected}{$i$} \label{alg:naive--isaff-begin}
    \Return{$1$}
  \EndFunction \label{alg:naive--isaff-end}
  \Function{inAffectedRange}{$i$} \label{alg:naive--isaffrng-begin}
    \Return{$1$}
  \EndFunction \label{alg:naive--isaffrng-end}
  \State $F \gets \{isAffected, inAffectedRange\}$ \label{alg:naive--lambdas}
  \State $\rhd$ Use $K^{t-1}$, $\Sigma^{t-1}$ as auxiliary information (Alg. \ref{alg:update})
  \State $\{K^t, \Sigma^t\} \gets updateWeights(G^t, \Delta^{t-}, \Delta^{t+}, C^{t-1}, K^{t-1}, \Sigma^{t-1})$\label{alg:naive--auxliliary}
  \State $\rhd$ Obtain updated communities (Alg. \ref{alg:leiden})
  \State $C^t \gets leiden(G^t, C^{t-1}, K^t, \Sigma^t, F)$ \label{alg:naive--leiden}
  \Return{$\{C^t, K^t, \Sigma^t\}$} \label{alg:naive--return}
\EndFunction
\end{algorithmic}
\end{algorithm}

%% file: src/alg-delta.tex
\begin{algorithm}[hbtp]
\caption{Our Parallel \textit{Delta-screening (DS)} Leiden \cite{sahu2024dflouvain}.}
\label{alg:delta}
\begin{algorithmic}[1]
\Require{$G^t(V^t, E^t)$: Current/updated input graph}
\Require{$\Delta^{t-}, \Delta^{t+}$: Edge deletions and insertions (batch update)}
\Require{$C^{t-1}, C^t$: Previous, current community of each vertex}
\Require{$K^{t-1}, K^t$: Previous, current weighted-degree of vertices}
\Require{$\Sigma^{t-1}, \Sigma^t$: Previous, current total edge weight of communities}
\Ensure{$\delta V, \delta E, \delta C$: Is vertex, neighbors, or community affected?}
\Ensure{$H$: Hashtable mapping a community to associated weight}
\Ensure{$isAffected(i)$: Is vertex $i$ is marked as affected?}
\Ensure{$inAffectedRange(i)$: Can $i$ be incrementally marked?}
\Ensure{$F$: Lambda functions passed to parallel Leiden (Alg. \ref{alg:leiden})}

\Statex

\Function{deltaScreening}{$G^t, \Delta^{t-}, \Delta^{t+}, C^{t-1}, K^{t-1}, \Sigma^{t-1}$}
  \State $H, \delta V, \delta E, \delta C \gets \{\}$ \label{alg:delta--init}
  \State $\rhd$ Mark affected vertices
  \ForAll{$(i, j, w) \in \Delta^{t-}$ \textbf{in parallel}} \label{alg:delta--loopdel-begin}
    \If{$C^{t-1}[i] = C^{t-1}[j]$}
      \State $\delta V[i], \delta E[i], \delta C[C^{t-1}[j]] \gets 1$ \label{alg:delta--loopdelmark}
    \EndIf
  \EndFor \label{alg:delta--loopdel-end}
  \ForAll{unique source vertex $i \in \Delta^{t+}$ \textbf{in parallel}} \label{alg:delta--loopins-begin}
    \State $H \gets \{\}$
    \ForAll{$(i', j, w) \in \Delta^{t+}\ |\ i' = i$} \label{alg:delta--loopinssrc-begin}
      \If{$C^{t-1}[i] \neq C^{t-1}[j]$}
        \State $H[C^{t-1}[j]] \gets H[C^{t-1}[j]] + w$
      \EndIf
    \EndFor \label{alg:delta--loopinssrc-end}
    \State $c^* \gets$ Best community linked to $i$ in $H$ \label{alg:delta--loopinschoose}
    \State $\delta V[i], \delta E[i], \delta C[c^*] \gets 1$ \label{alg:delta--loopinsmark}
  \EndFor \label{alg:delta--loopins-end}
  \ForAll{$i \in V^t$ \textbf{in parallel}} \label{alg:delta--loopaff-begin}
    \If{$\delta E[i]$} \label{alg:delta--loopaffnei-begin}
      \ForAll{$j \in G^t.neighbors(i)$}
        \State $\delta V[j] \gets 1$
      \EndFor
    \EndIf \label{alg:delta--loopaffnei-end}
    \If{$\delta C[C^{t-1}[i]]$} \label{alg:delta--loopaffcom-begin}
      \State $\delta V[i] \gets 1$
    \EndIf \label{alg:delta--loopaffcom-end}
  \EndFor \label{alg:delta--loopaff-end}
  \Function{isAffected}{$i$} \label{alg:delta--isaff-begin}
    \Return{$\delta V[i]$}
  \EndFunction \label{alg:delta--isaff-end}
  \Function{inAffectedRange}{$i$} \label{alg:delta--isaffrng-begin}
    \Return{$\delta V[i]$}
  \EndFunction \label{alg:delta--isaffrng-end}
  \State $F \gets \{isAffected, inAffectedRange\}$ \label{alg:delta--lambdas}
  \State $\rhd$ Use $K^{t-1}$, $\Sigma^{t-1}$ as auxiliary information (Alg. \ref{alg:update})
  \State $\{K^t, \Sigma^t\} \gets updateWeights(G^t, \Delta^{t-}, \Delta^{t+}, C^{t-1}, K^{t-1}, \Sigma^{t-1})$\label{alg:delta--auxiliary}
  \State $\rhd$ Obtain updated communities (Alg. \ref{alg:leiden})
  \State $C^t \gets leiden(G^t, C^{t-1}, K^t, \Sigma^t, F)$ \label{alg:delta--leiden}
  \Return{$\{C^t, K^t, \Sigma^t\}$} \label{alg:delta--return}
\EndFunction
\end{algorithmic}
\end{algorithm}

%% file: src/alg-leidenlm.tex
\begin{algorithm}[hbtp]
\caption{Local-moving phase of our Parallel Leiden \cite{sahu2024fast}.}
\label{alg:leidenlm}
\begin{algorithmic}[1]
\Require{$G'(V', E')$: Input/super-vertex graph}
\Require{$C'$: Community membership of each vertex}
\Require{$K'$: Total edge weight of each vertex}
\Require{$\Sigma'$: Total edge weight of each community}
\Require{$\Delta C'$: Changed communities flag}
\Require{$F$: Lambda functions passed to parallel Leiden}
\Ensure{$H_t$: Collision-free per-thread hashtable}
\Ensure{$l_i$: Number of iterations performed}
\Ensure{$\tau$: Per iteration tolerance}

\Statex

\Function{leidenMove}{$G', C', K', \Sigma', \Delta C', F$} \label{alg:leidenlm--move-begin}
  \ForAll{$l_i \in [0 .. \text{MAX\_ITERATIONS})$} \label{alg:leidenlm--iterations-begin}
    \State Total delta-modularity per iteration: $\Delta Q \gets 0$ \label{alg:leidenlm--init-deltaq}
    \ForAll{unprocessed $i \in V'$ \textbf{in parallel}} \label{alg:leidenlm--loop-vertices-begin}
      \State Mark $i$ as processed (prune) \label{alg:leidenlm--prune}
      \If{\textbf{not} $F.inAffectedRange(i)$} \textbf{continue} \label{alg:leidenlm--affrng}
      \EndIf
      \State $H_t \gets scanCommunities(\{\}, G', C', i, false)$ \label{alg:leidenlm--scan}
      \State $\rhd$ Use $H_t, K', \Sigma'$ to choose best community
      \State $c^* \gets$ Best community linked to $i$ in $G'$ \label{alg:leidenlm--best-community-begin}
      \State $\delta Q^* \gets$ Delta-modularity of moving $i$ to $c^*$ \label{alg:leidenlm--best-community-end}
      \If{$c^* = C'[i]$} \textbf{continue} \label{alg:leidenlm--best-community-same}
      \EndIf
      \State $\Sigma'[C'[i]] -= K'[i]$ \textbf{;} $\Sigma'[c^*] += K'[i]$ \textbf{atomic} \label{alg:leidenlm--perform-move-begin}
      \State $C'[i] \gets c^*$ \textbf{;} $\Delta Q \gets \Delta Q + \delta Q^*$ \label{alg:leidenlm--perform-move-end}
      \State Mark neighbors of $i$ as unprocessed \label{alg:leidenlm--remark}
      \If{is dynamic alg.} $\Delta C'[c] \gets \Delta C'[c^*] \gets 1$ \label{alg:leidenlm--mark-changed-communities}
      \EndIf
    \EndFor \label{alg:leidenlm--loop-vertices-end}
    \If{$\Delta Q \le \tau$} \textbf{break} \Comment{Locally converged?} \label{alg:leidenlm--locally-converged}
    \EndIf
  \EndFor \label{alg:leidenlm--iterations-end}
  \Return{$l_i$} \label{alg:leidenlm--return}
\EndFunction \label{alg:leidenlm--move-end}

\Statex

\Function{scanCommunities}{$H_t, G', C', i, self$}
  \ForAll{$(j, w) \in G'.edges(i)$}
    \If{\textbf{not} $self$ \textbf{and} $i = j$} \textbf{continue}
    \EndIf
    \State $H_t[C'[j]] \gets H_t[C'[j]] + w$
  \EndFor
  \Return{$H_t$}
\EndFunction
\end{algorithmic}
\end{algorithm}


%% file: src/alg-leidenre.tex
\begin{algorithm}[hbtp]
\caption{Refinement phase of our Parallel Leiden \cite{sahu2024fast}.}
\label{alg:leidenre}
\begin{algorithmic}[1]
\Require{$G'(V', E')$: Input/super-vertex graph}
\Require{$C'_B$: Community bound of each vertex}
\Require{$C'$: Community membership of each vertex}
\Require{$K'$: Total edge weight of each vertex}
\Require{$\Sigma'$: Total edge weight of each community}
\Require{$\Delta C'$: Community changed flag}
\Ensure{$G'_{C'}$: Community vertices (CSR)}
\Ensure{$H_t$: Collision-free per-thread hashtable}
\Ensure{$\tau$: Per iteration tolerance}

\Statex

\Function{leidenRefine}{$G', C'_B, C', K', \Sigma', \Delta C', \tau$} \label{alg:leidenre--move-begin}
  \ForAll{$i \in V'$ \textbf{in parallel}} \label{alg:leidenre--loop-vertices-begin}
    \State $c \gets C'[i]$
    \If{$\Delta C'[c] = 0$ \textbf{or} $\Sigma'[c] \neq K'[i]$} \textbf{continue} \label{alg:leidenre--check-isolated}
    \EndIf
    \State $H_t \gets scanBounded(\{\}, G', C'_B, C', i, false)$ \label{alg:leidenre--scan}
    \State $\rhd$ Use $H_t, K', \Sigma'$ to choose best community
    \State $c^* \gets$ Best community linked to $i$ in $G'$ within $C'_B$ \label{alg:leidenre--best-community-begin}
    \State $\delta Q^* \gets$ Delta-modularity of moving $i$ to $c^*$ \label{alg:leidenre--best-community-end}
    \If{$c^* = c$ \textbf{or} $C'[c^*] \neq c^* $} \textbf{continue} \label{alg:leidenre--best-community-same}
    \EndIf
    \If{$atomicCAS(\Sigma'[c], K'[i], 0) = K'[i]$} \label{alg:leidenre--perform-move-begin}
      \State $\Sigma'[c^*] += K'[i]$ \textbf{atomically}
      \State $C'[i] \gets c^*$ \label{alg:leidenre--perform-move-end}
    \EndIf
  \EndFor \label{alg:leidenre--loop-vertices-end}
\EndFunction \label{alg:leidenre--move-end}

\Statex

\Function{scanBounded}{$H_t, G', C'_B, C', i, self$}
  \ForAll{$(j, w) \in G'.edges(i)$}
    \If{\textbf{not} $self$ \textbf{and} $i = j$} \textbf{continue}
    \EndIf
    \If{$C'_B[i] \neq C'_B[j]$} \textbf{continue}
    \EndIf
    \State $H_t[C'[j]] \gets H_t[C'[j]] + w$
  \EndFor
  \Return{$H_t$}
\EndFunction

\Statex

\Function{atomicCAS}{$pointer, old, new$}
  \State $\rhd$ Perform the following atomically
  \If{$pointer = old$} $pointer \gets new$ \textbf{;} \ReturnInline{$old$}
  \Else\ \ReturnInline{$pointer$}
  \EndIf
\EndFunction
\end{algorithmic}
\end{algorithm}

%% file: src/alg-leidenag.tex
\begin{algorithm}[hbtp]
\caption{Aggregation phase of our Parallel Leiden \cite{sahu2024fast}.}
\label{alg:leidenag}
\begin{algorithmic}[1]
\Require{$G'(V', E')$: Input/super-vertex graph}
\Require{$C'$: Community membership of each vertex}
\Ensure{$G'_{C'}$: Community vertices (CSR)}
\Ensure{$G''$: Super-vertex graph (weighted CSR)}
\Ensure{$*.offsets$: Offsets array of a CSR graph}
\Ensure{$H_t$: Collision-free per-thread hashtable}

\Statex

\Function{leidenAggregate}{$G', C'$}
  \State $\rhd$ Obtain vertices belonging to each community
  \State $G'_{C'}.offsets \gets countCommunityVertices(G', C')$ \label{alg:leidenag--coff-begin}
  \State $G'_{C'}.offsets \gets exclusiveScan(G'_{C'}.offsets)$ \label{alg:leidenag--coff-end}
  \ForAll{$i \in V'$ \textbf{in parallel}} \label{alg:leidenag--comv-begin}
    \State Add edge $(C'[i], i)$ to CSR $G'_{C'}$ atomically
  \EndFor \label{alg:leidenag--comv-end}
  \State $\rhd$ Obtain super-vertex graph
  \State $G''.offsets \gets communityTotalDegree(G', C')$ \label{alg:leidenag--yoff-begin}
  \State $G''.offsets \gets exclusiveScan(G''.offsets)$ \label{alg:leidenag--yoff-end}
  \State $|\Gamma| \gets$ Number of communities in $C'$
  \ForAll{$c \in [0, |\Gamma|)$ \textbf{in parallel}} \label{alg:leidenag--y-begin}
    \If{degree of $c$ in $G'_{C'} = 0$} \textbf{continue}
    \EndIf
    \State $H_t \gets \{\}$
    \ForAll{$i \in G'_{C'}.edges(c)$}
      \State $H_t \gets scanCommunities(H, G', C', i, true)$
    \EndFor
    \ForAll{$(d, w) \in H_t$}
      \State Add edge $(c, d, w)$ to CSR $G''$ atomically
    \EndFor
  \EndFor \label{alg:leidenag--y-end}
  \Return $G''$ \label{alg:leidenag--return}
\EndFunction
\end{algorithmic}
\end{algorithm}

%% file: src/alg-update.tex
\begin{algorithm}[hbtp]
\caption{Updating vertex/community weights in parallel\ignore{\cite{sahu2024dflouvain}}.}
\label{alg:update}
\begin{algorithmic}[1]
\Require{$G^t(V^t, E^t)$: Current input graph}
\Require{$\Delta^{t-}, \Delta^{t+}$: Edge deletions and insertions (batch update)}
\Require{$C^{t-1}$: Previous community of each vertex}
\Require{$K^{t-1}$: Previous weighted-degree of each vertex}
\Require{$\Sigma^{t-1}$: Previous total edge weight of each community}
\Ensure{$K$: Updated weighted-degree of each vertex}
\Ensure{$\Sigma$: Updated total edge weight of each community}
\Ensure{$work_{th}$: Work-list of current thread}

\Statex

\Function{updateWeights}{$G^t, \Delta^{t-}, \Delta^{t+}, C^{t-1}, K^{t-1}, \Sigma^{t-1}$}
  \State $K \gets K^{t-1}$ \textbf{;} $\Sigma \gets \Sigma^{t-1}$ \label{alg:update--init}
  \ForAll{\textbf{threads in parallel}} \label{alg:update--loopdel-begin}
    \ForAll{$(i, j, w) \in \Delta^{t-}$}
      \State $c \gets C^{t-1}[i]$ \label{alg:update--delc}
      \If{$i \in work_{th}$} $K[i] \gets K[i] - w$ \label{alg:update--delk}
      \EndIf
      \If{$c \in work_{th}$} $\Sigma[c] \gets \Sigma[c] - w$ \label{alg:update--delsigma}
      \EndIf
    \EndFor \label{alg:update--loopdel-end}
    \ForAll{$(i, j, w) \in \Delta^{t+}$} \label{alg:update--loopins-begin}
      \State $c \gets C^{t-1}[i]$
      \If{$i \in work_{th}$} $K[i] \gets K[i] + w$
      \EndIf
      \If{$c \in work_{th}$} $\Sigma[c] \gets \Sigma[c] + w$
      \EndIf
    \EndFor
  \EndFor \label{alg:update--loopins-end}
  \Return $\{K, \Sigma\}$ \label{alg:update--return}
\EndFunction
\end{algorithmic}
\end{algorithm}

%% file: src/fig-temporal-summary.tex
\begin{figure*}[!hbt]
  \centering
  \subfigure[Overall Runtime]{
    \label{fig:temporal-summary--runtime-overall}
    \includegraphics[width=0.48\linewidth]{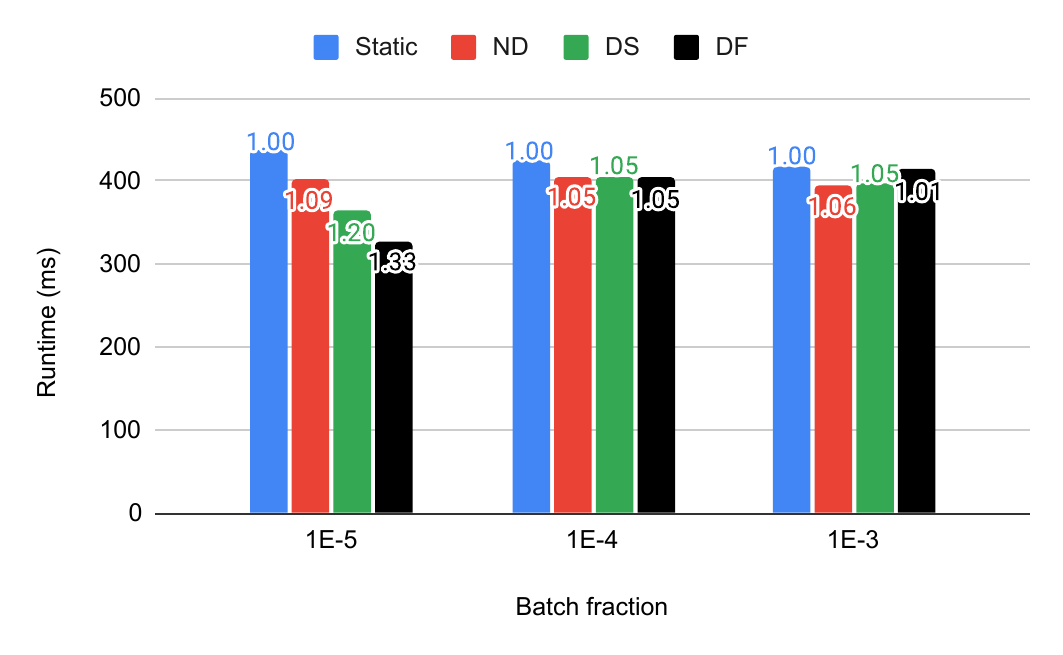}
  }
  \subfigure[Overall Modularity of communities obtained]{
    \label{fig:temporal-summary--modularity-overall}
    \includegraphics[width=0.48\linewidth]{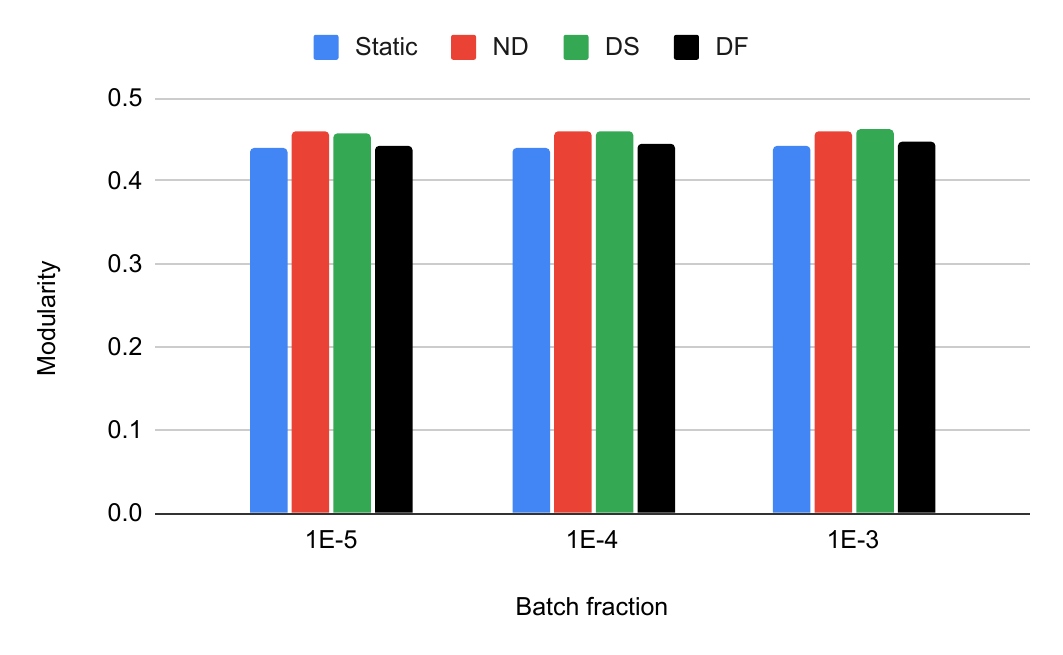}
  } \\[2ex]
  \includegraphics[width=0.48\linewidth]{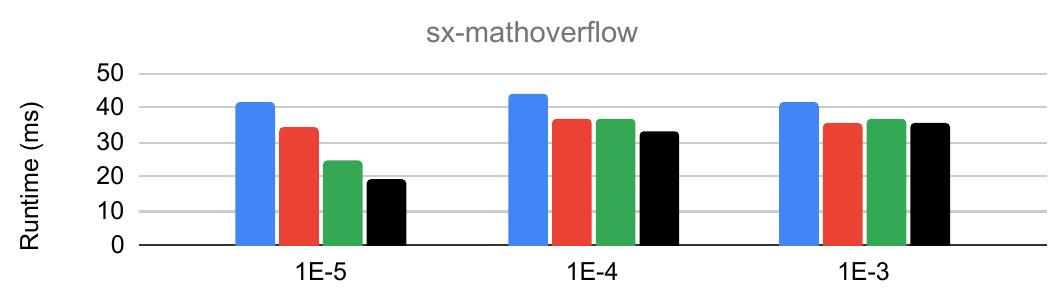}
  \includegraphics[width=0.48\linewidth]{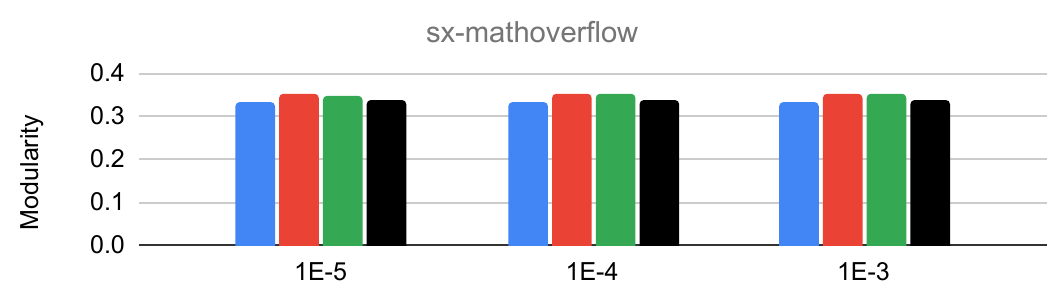}
  \includegraphics[width=0.48\linewidth]{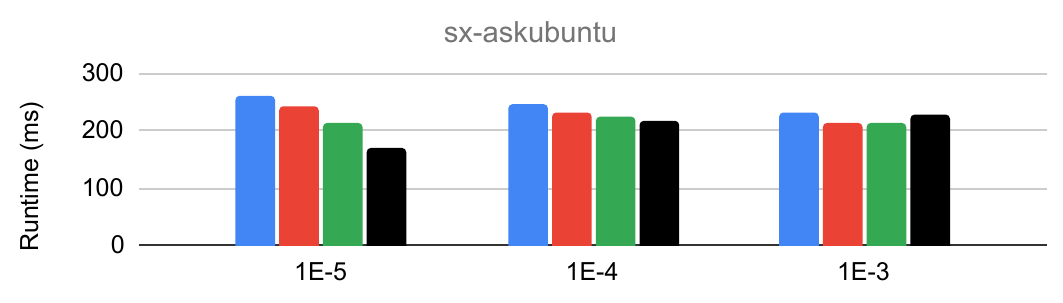}
  \includegraphics[width=0.48\linewidth]{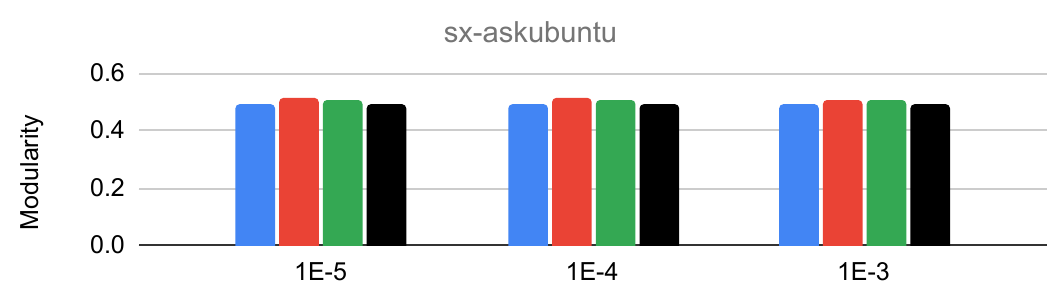}
  \includegraphics[width=0.48\linewidth]{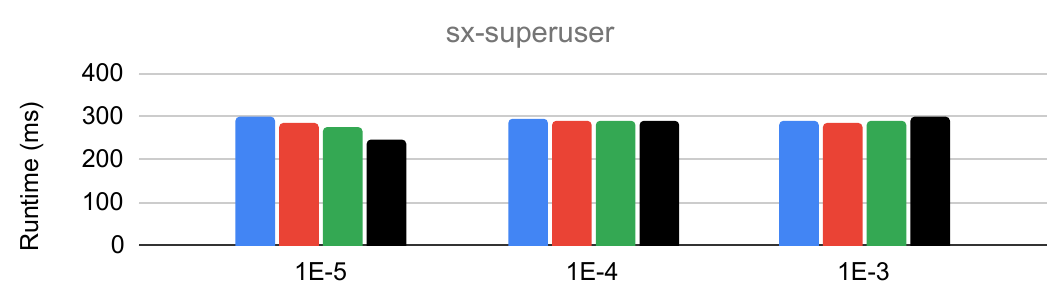}
  \includegraphics[width=0.48\linewidth]{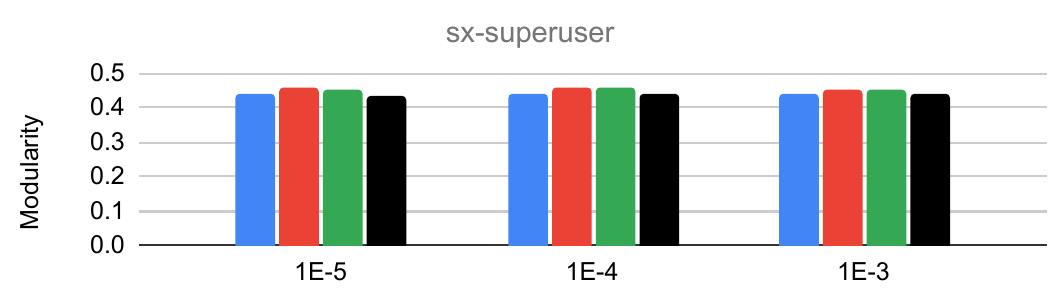}
  \includegraphics[width=0.48\linewidth]{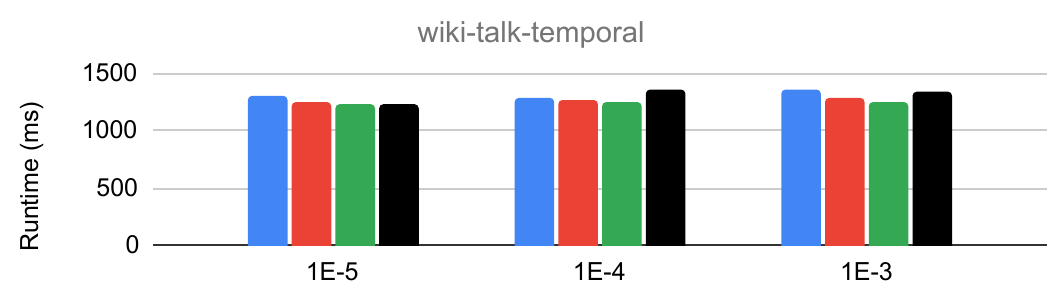}
  \includegraphics[width=0.48\linewidth]{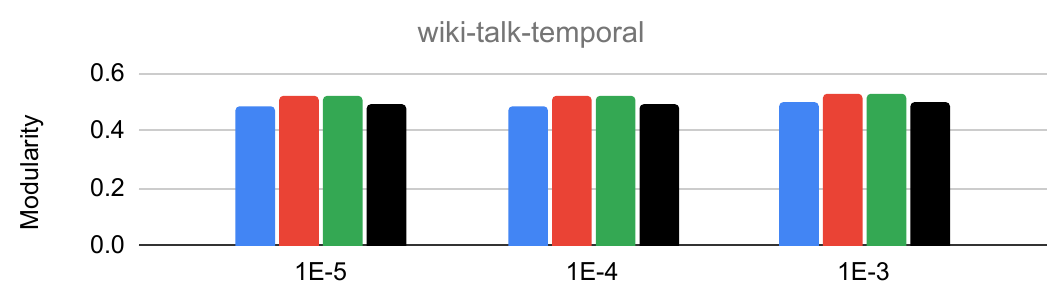}
  \subfigure[Runtime on each dynamic graph]{
    \label{fig:temporal-summary--runtime-graph}
    \includegraphics[width=0.48\linewidth]{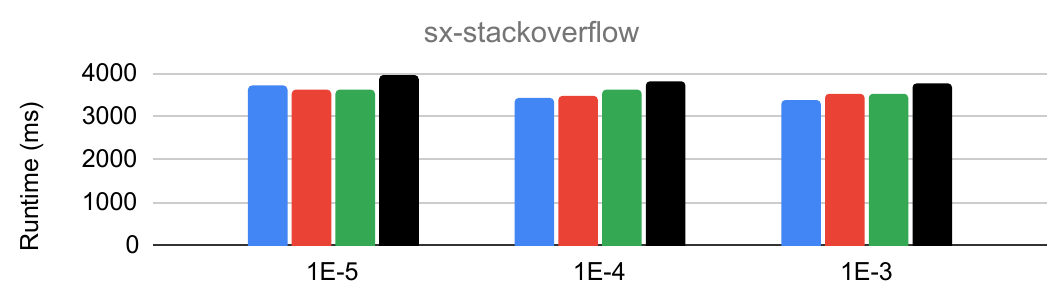}
  }
  \subfigure[Modularity in communities obtained on each dynamic graph]{
    \label{fig:temporal-summary--modularity-graph}
    \includegraphics[width=0.48\linewidth]{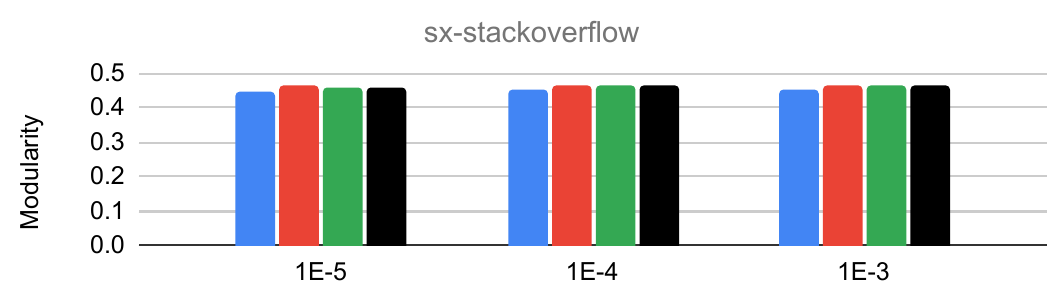}
  } \\[-2ex]
  \caption{Mean Runtime and Modularity of communities obtained with our multicore implementation of \textit{Static}, \textit{Naive-dynamic (ND)}, \textit{Delta-screening (DS)}, and \textit{Dynamic Frontier (DF)} Leiden on real-world dynamic graphs, using batch updates of size $10^{-5}|E_T|$ to $10^{-3}|E_T|$. Here, (a) and (b) display the overall runtime and modularity across all temporal graphs, while (c) and (d) display the runtime and modularity for each individual graph. In (a), the speedup of each approach relative to Static Leiden is labeled.}
  \label{fig:temporal-summary}
\end{figure*}